\documentclass[twocolumn]{aastex701}
\usepackage{graphicx}
\usepackage{amsmath}
\usepackage{hyperref}

\begin{document}

\title{\Large Strong MHD Turbulence and Coherent Structures as Cosmic Particle Accelerators}

\author[orcid=0000-0002-8700-4172,sname='North America']{Loukas Vlahos}
\altaffiliation{Department of Physics, University of Ioannina, 11110 Ioannina, Greece}
\affiliation{Department of Physics, Aristotle  University, 54124 Thessaloniki, Greece}
\email[show]{vlahos@astro.auth.gr}  

\begin{abstract}
Magnetohydrodynamic (MHD) turbulence is a ubiquitous dynamical state of astrophysical plasmas and a primary agent in the redistribution, dissipation, and conversion of energy into particle populations. Yet turbulence is still most often described in terms of cascades, spectra, and scale-to-scale transfer, while its role in producing localized sites of intense energization remains comparatively underemphasized. In this forward-looking review, aimed at a broad astrophysical readership, I argue that any physically complete picture of turbulent plasma heating and particle acceleration must place the self-consistent emergence of coherent structures at its center. Current sheets, vortical structures, magnetic flux ropes, shocklets, and confined reconnection sites are not secondary by-products of the turbulent cascade; they are its dynamically dominant dissipative and energizing elements, where electric fields intensify, dissipation becomes highly localized, and particles undergo repeated acceleration.

Viewed in this way, strong turbulence provides a unifying framework that links large-scale plasma dynamics to the generation of suprathermal particles and non-thermal energy distributions in the solar atmosphere, the solar wind, shock environments, and a wide range of other cosmic plasmas. Rather than attempting an exhaustive survey of the literature, this article offers a selective and physically organized synthesis of the field, emphasizing the mechanisms, regimes, and open problems most relevant to the development of predictive theories of particle acceleration in turbulent plasmas. It also identifies the principal conceptual and computational challenges that must be overcome if the next generation of models is to connect multiscale plasma dynamics with observable energetic-particle signatures.
\end{abstract}

\keywords{MHD turbulence, coherent structure formation, energetic particles, high energy astrophysics, cosmic rays}


\section{Introduction}

In astrophysics, turbulence is commonly described within the classical framework of the Reynolds number, the transfer of energy across spatial scales, and the Kolmogorov picture of hydrodynamic turbulence. Although these concepts remain foundational, most astrophysical plasmas are magnetized, only weakly collisional, and far from thermodynamic equilibrium. Under such conditions, the coupling between velocity and magnetic fields fundamentally alters the dynamics. Magnetohydrodynamic (MHD) turbulence is therefore not simply a magnetic extension of ordinary fluid turbulence, but a richer multiscale process shaped by magnetic tension, anisotropy, intermittency, magnetic field-line topology, and reconnection \citep{Matthaeus11,MatthaeusREv2021,Vlahos23}.

As turbulence develops, the plasma self-organizes into localized coherent structures. Within these regions, gradients steepen, electric fields intensify, dissipation becomes concentrated, and particles undergo trapping, scattering, and repeated episodes of acceleration. These structures therefore provide the physical link between large-scale turbulent dynamics and the small-scale processes responsible for plasma heating, the formation of suprathermal particle populations, and the emergence of non-thermal energy distributions.

This viewpoint also changes how particle acceleration in astrophysical plasmas should be interpreted. Mechanisms that are often studied separately—stochastic acceleration, shock acceleration, reconnection-driven acceleration, and wave--particle interactions—may in fact operate simultaneously within the same intermittent turbulent environment. From this perspective, strong MHD turbulence offers a unifying framework in which these apparently distinct energization channels arise as interconnected consequences of multiscale energy transfer in a spatially structured electromagnetic field.

\paragraph*{Scope and perspective:} \, This article is not intended to be a comprehensive review. Rather, it presents a forward-looking and concise overview with a pedagogical emphasis, aimed at astrophysicists who are not specialists in turbulence. The focus is on how coherent structures emerge, how they modify electric fields, how these processes transfer energy across scales, and how they influence particle acceleration in a broad range of magnetized plasma environments.

The presentation is organized as follows. I begin by outlining the transition from ordinary fluid turbulence to MHD turbulence, emphasizing the features most relevant to particle energization. The next section examines the emergence of coherent structures in strongly turbulent MHD regimes and discusses their statistical properties. I then explore how such coherent structures manifest themselves in laboratory, space, and astrophysical plasmas. Finally, I review numerical approaches—including MHD simulations, test-particle methods, and kinetic techniques—and discuss future prospects for the development of reduced, physics-informed neural models of turbulent particle acceleration.

\section{From Hydrodynamic to Magnetohydrodynamic Turbulence}

\begin{figure}[!ht]
    \centering
\includegraphics[width=0.5\linewidth]{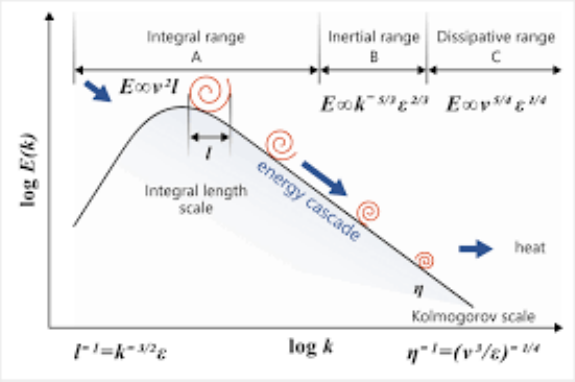}
    \caption{Energy introduced at large scales is gradually transferred to progressively smaller scales until it reaches the dissipation range, where it is converted into heat.}
    \label{fig:Kolmog1}
\end{figure}

A standard way to introduce turbulence is through the competition between nonlinear advection and dissipation. In an ordinary fluid, this balance is measured by the Reynolds number,
\begin{equation}
Re = \frac{uL}{\nu},
\end{equation}
where $u$ is a characteristic flow speed, $L$ is a representative large length scale, and $\nu$ is the kinematic viscosity. When $Re \gg 1$, nonlinear advection dominates over viscous diffusion, the flow becomes unstable, and eddies develop over a broad hierarchy of scales. In this regime, energy injected at large scales is progressively transferred to smaller scales until it reaches the dissipative range, where it is ultimately converted into heat (Fig. \ref{fig:Kolmog1}). This cascade picture, established by Kolmogorov, remains one of the basic organizing principles of turbulence theory \citep{Kolmogorov1941a, Kolmogorov1941b}.

Within the inertial range of fully developed hydrodynamic turbulence, the mean energy flux per unit mass, $\epsilon$, is approximately constant across scales. For an eddy of size $\ell$ and characteristic velocity $u_\ell$, the turnover time is $\tau_\ell \sim \ell/u_\ell$, while the energy transfer rate is
\begin{equation}
\Pi_\ell \sim \frac{u_\ell^2}{\tau_\ell} \sim \frac{u_\ell^3}{\ell}.
\end{equation}
By setting $\Pi_\ell \sim \epsilon$, one recovers the Kolmogorov scaling
\begin{equation}
u_\ell \sim (\epsilon \ell)^{1/3},
\end{equation}
which implies the familiar energy spectrum
\begin{equation}
E(k) \propto \epsilon^{2/3} k^{-5/3}.
\end{equation}
Although idealized, this framework captures the essential point that turbulence is intrinsically multiscale, with nonlinear interactions transporting energy across scales in an approximately self-similar way.

Astrophysical plasmas, however, are rarely well described as neutral fluids. In most space and astrophysical environments, magnetic fields are dynamically important, and the appropriate macroscopic description is magnetohydrodynamics (MHD). The dynamics are then controlled not only by inertia, pressure gradients, and viscous stresses, but also by magnetic pressure, magnetic tension, and the evolution of the magnetic field itself. The MHD momentum and induction equations may be written as \citep{Boyd69, Krall73}
\begin{equation}
\rho \frac{d\mathbf{u}}{dt}
=
-\nabla \left(P + \frac{B^2}{8\pi}\right)
+
\frac{(\mathbf{B}\cdot\nabla)\mathbf{B}}{4\pi}
+
\nu \rho \nabla^2 \mathbf{u}
+
\mathbf{F},
\end{equation}
\begin{equation}
\frac{\partial \mathbf{B}}{\partial t}
=
\nabla \times (\mathbf{u}\times\mathbf{B})
+
\eta \nabla^2 \mathbf{B},
\end{equation}
where $\rho$ is the plasma density, $\mathbf{u}$ the bulk velocity, $\mathbf{B}$ the magnetic field, $\nu$ the kinematic viscosity, and $\eta$ the magnetic diffusivity. The induction equation expresses a defining property of plasma dynamics: the magnetic field is not a passive tracer. It is advected, stretched, and diffused by the flow, while simultaneously reacting back on the plasma through the Lorentz force.

The corresponding control parameter for magnetic evolution is the magnetic Reynolds number,
\begin{equation}
R_m = \frac{uL}{\eta},
\end{equation}
which measures the competition between inductive transport and magnetic diffusion. In typical astrophysical plasmas, both $Re$ and $R_m$ are extremely large, implying that velocity and magnetic fluctuations evolve nonlinearly over an enormous range of scales. Under these conditions, the cascade becomes intrinsically coupled: kinetic and magnetic energies interact continuously, producing anisotropy, intermittency, and strong spatial inhomogeneity. This already represents a major departure from the simplest hydrodynamic picture.

A further distinction concerns the nature of the cascade itself. In magnetized plasmas, turbulent fluctuations can propagate as Alfv\'enic disturbances, are shaped by the local magnetic geometry, and often cascade anisotropically with respect to the mean or local magnetic field. Just as importantly, nonlinear MHD dynamics do not merely generate a background of random fluctuations; they also produce sharp gradients and complex magnetic topology. MHD turbulence, therefore, acts not only as a cascade process but also as an efficient generator of structure.

This point is central to particle energization. The quantity that enters directly into the dynamics of charged particles is the electric field, which in MHD is given by
\begin{equation}
\mathbf{E}
=
-\frac{\mathbf{u}\times\mathbf{B}}{c}
+
\eta \mathbf{J},
\end{equation}
with
\begin{equation}
\mathbf{J} = \frac{c}{4\pi}\nabla\times\mathbf{B}.
\end{equation}
The electric field, therefore, depends precisely on the quantities that become highly intermittent in strong turbulence: the velocity, the magnetic field, and the current density. Once the plasma enters a regime of large magnetic fluctuations and abundant coherent structures, the electromagnetic environment experienced by particles becomes strongly inhomogeneous in both space and time. It is then natural to expect that transport and acceleration are governed not only by broadband fluctuations, but also by repeated interactions with localized structures where electric fields are enhanced, and dissipation is concentrated \citep{Vlahos04}.

For the purposes of this review, this is the essential conceptual step. Hydrodynamic turbulence shows how nonlinear systems redistribute energy across scales. MHD turbulence adds a further level of complexity by coupling this transfer to magnetic topology, intermittent current structures, and spatially localized electric fields. It is precisely this combination that makes strong MHD turbulence so relevant to astrophysical plasmas: it provides not only a means of redistributing energy, but a genuinely dynamical environment in which plasma heating, reconnection, and particle acceleration can be understood within a unified multiscale framework.

\section{The Emergence of Coherent Structures in Strong MHD Turbulence}

Intermittency is one of the defining signatures of strong turbulence. In a turbulent plasma, dynamical activity is distributed unevenly: extended regions may remain relatively quiet, whereas only a limited fraction of the volume contains sharp gradients, intense currents, enhanced vorticity, and strong electric fields. In this way, intermittency links the large-scale transfer of energy across scales to the localized regions where that energy is ultimately converted into heat or non-thermal particle energy.

Several studies \citep{Dmitruk03, Dmitruk04, Arzner06} investigated decaying turbulence in simulations that started with both kinetic and magnetic perturbations, considering configurations both with and without an imposed magnetic field, as well as setups featuring a strong guide field \citep{Dong22}.
\begin{figure}[!ht]
    \centering
    \includegraphics[width=0.45\linewidth]{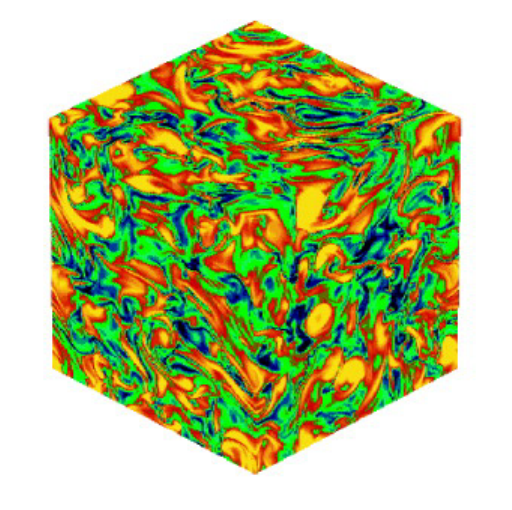}
    \includegraphics[width=0.45\linewidth]{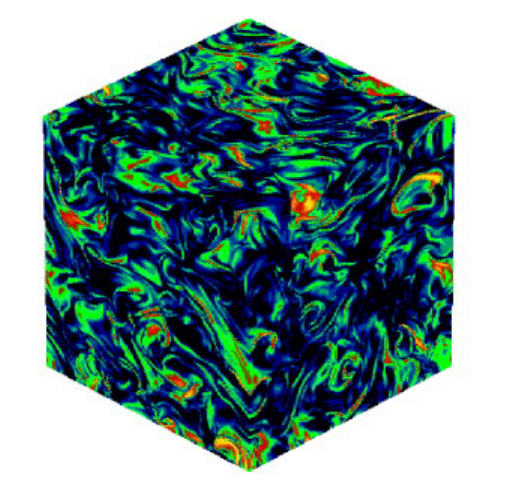}
    \caption{Visualization of the turbulent magnetic field strength $|B|$ (left) and electric field strength $|\mathbf{E}|$ (right) across the simulation domain. Regions of high intensity are shown in yellow (light), whereas low-intensity regions are shown in blue (dark) \citep{Dmitruk03}.}
    \label{fig:Dmitruk03}
\end{figure}

After several Alfv\'en crossing times, $\tau_A$ $(\tau_A = L/v_A$, with $v_A$ denoting the Alfv\'en speed \citep{ChenBook16}), the system reaches a fully developed turbulent state spanning a broad range of spatial scales (see Fig.~\ref{fig:Dmitruk03}). In this regime, coherent structures associated with intense, spatially localized electric fields naturally arise from the nonlinear interaction of kinetic and magnetic fluctuations.

\begin{figure}[!ht]
    \centering
    \includegraphics[width=0.3\linewidth]{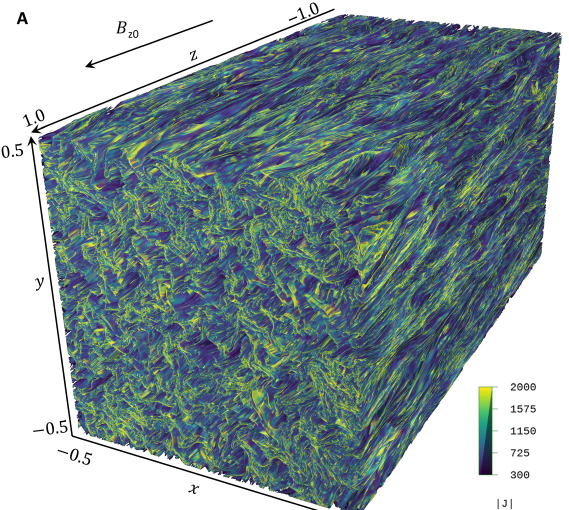}
    \includegraphics[width=0.35\linewidth]{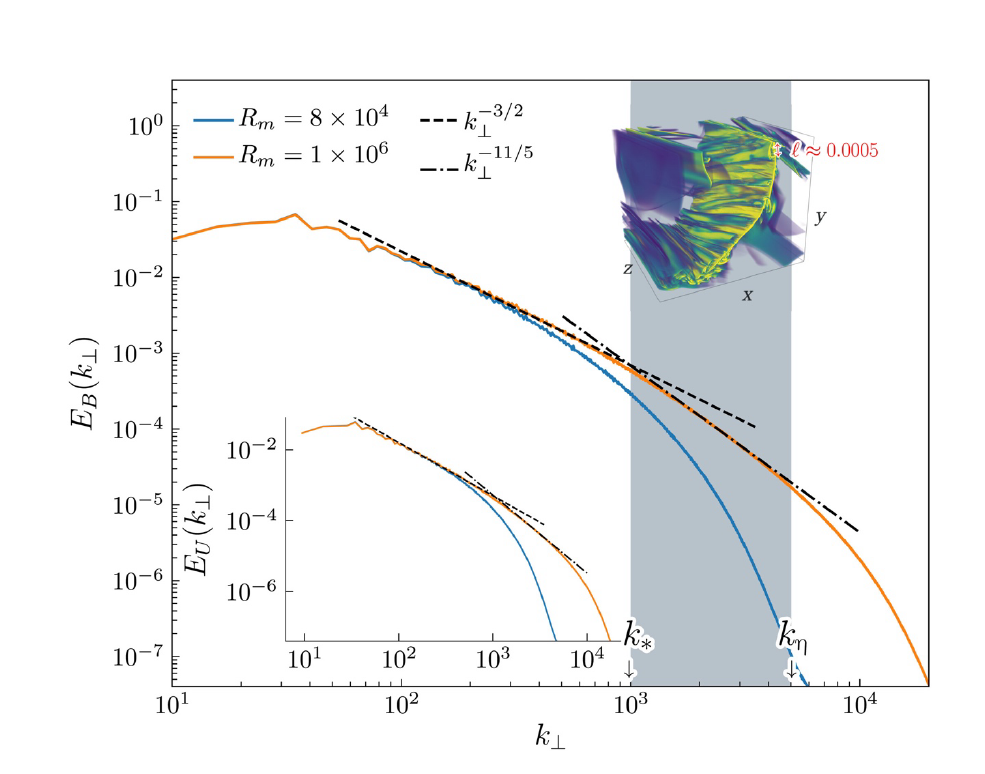}
   
    \caption{(Left) Volume rendering of the current density $\lvert J \rvert$ over the full domain at a time when turbulence is fully developed. Numerous current sheets are visible in the plane perpendicular to the mean magnetic field $B_{z0}$. (Right) Spectra of the magnetic and kinetic energies associated with the perpendicular field components, $E_B(k_\perp)$ and $E_U(k_\perp)$, showing a standard inertial range with a slope $-3/2$
 and a reconnection-driven (or tearing-mediated) subinertial range with a slope $-11/5$
 for large $R_m$ (orange curves) \citep{Dong22}.}
    \label{fig:strongB}
\end{figure}

In strongly magnetized plasmas, where $B^2/4\pi > P$, the turbulent cascade becomes increasingly anisotropic and intermittent. Such conditions favor the formation of thin current sheets, magnetic filaments, vortices, and localized reconnection sites (see Fig. \ref{fig:strongB}) \citep{Dong22}.

Under such conditions, coherent structures are not uniformly distributed throughout space but instead appear in localized clusters. This patchy and highly inhomogeneous organization is consistent with a fractal or multifractal description, in which coherent structures occupy the plasma through a hierarchy of interconnected scales rather than as isolated entities with a single characteristic size \citep{Tu95, Isliker19}. Three-dimensional MHD simulations further show that both current and vorticity structures follow clear scaling laws in their spatial distributions, supporting the view that the topology of coherent structures is fundamentally fractal and self-similar over a wide range of scales \citep{Uritsky10}.

\begin{figure}[!ht]
    \centering
    \includegraphics[width=0.3\linewidth]{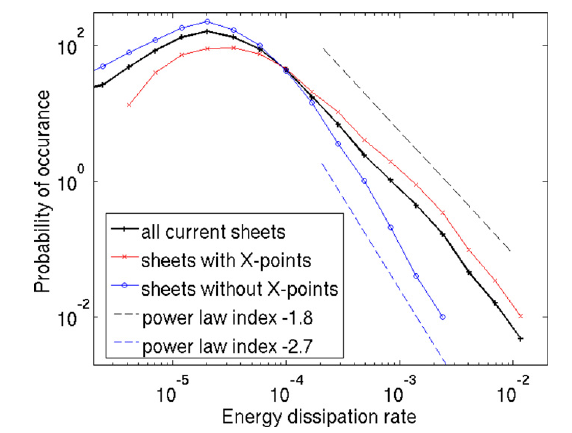}
     \includegraphics[width=0.3\linewidth]{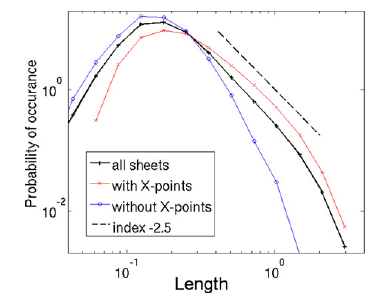}
    \caption{(Left) Probability distribution of the Ohmic energy dissipation rate in current sheets. The distribution for the full ensemble of current sheets (black) displays a power-law tail with an index of roughly $-1.8$. Also shown are the distributions for two subsets of current sheets: those containing X-points (red) and those without X-points (blue), the latter exhibiting a power-law index of approximately $-2.7$. (Right) Probability distributions for all geometric scales of the current sheets. For the full set of current sheets (black), the distributions of the length $l$ and width $\xi$ display power-law tails with exponents close to $-2.5$ \citep{Zhdankin13}.}
    \label{fig:Zh}
\end{figure}

\citet{Zhdankin13} analyzed fully developed MHD turbulence and provided a statistical characterization of the current sheets that arise spontaneously in their simulations. Their main result is that the characteristic properties of these structures follow power-law distributions. This is important because it shows that, in addition to the wave-like fluctuations cascading toward smaller scales, coherent structures in turbulent MHD also exhibit robust scale-free behavior (see Fig. \ref{fig:Zh}).

\begin{figure}[!ht]
    \centering
\includegraphics[width=0.4\linewidth]{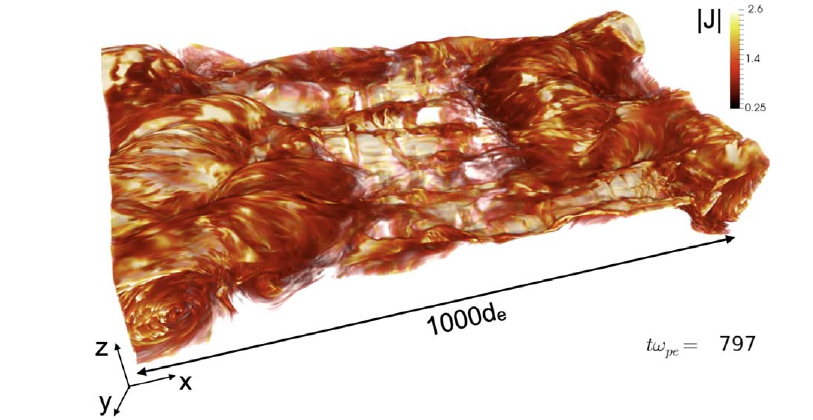}
\includegraphics[width=0.4\linewidth]{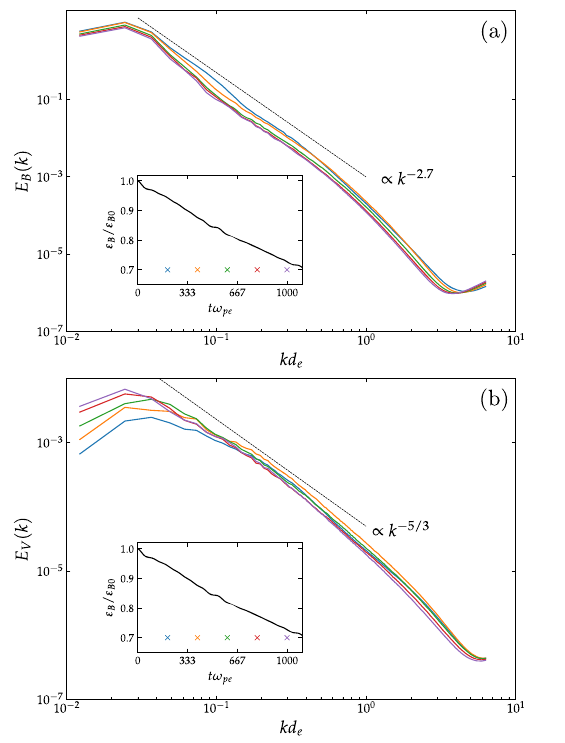}
    \caption{(Left) Turbulence generated by the fragmentation of a large-scale current sheet. (Right) Magnetic and kinetic energy spectra as functions of wavenumber at several times. The insets also show the temporal evolution of the system's total magnetic energy. For scales larger than the kinetic scale, the magnetic energy spectra follow an approximate power law with a slope of about $-2.7$, whereas the kinetic energy spectra are generally consistent with a $-5/3$ slope \citep{Guo21}.}
    \label{fig:FraqCS}
\end{figure}

\begin{figure}
    \centering
\includegraphics[width=0.7\linewidth]{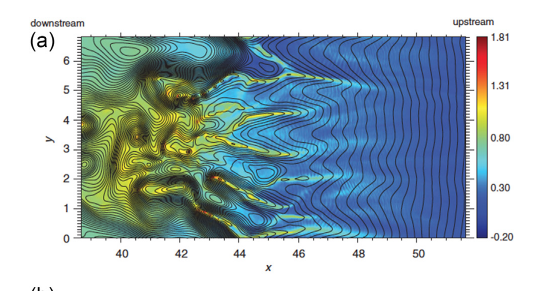}
    \caption{As time progresses, the evolution of a shock formed near the center of a turbulent medium produces an environment that becomes strongly turbulent \citep{Matsumoto15}.}
    \label{fig:MatsShck}
\end{figure}

It should also be emphasized that, in strongly turbulent regimes, coherent structures may arise either through the fragmentation of a large-scale current sheet \citep{Onofri06,Guo21,WangYulei25} (see Fig. \ref{fig:FraqCS}(Left)) or through the development of turbulence in the vicinity of a shock (see Fig. \ref{fig:MatsShck}) \citep{Matsumoto15, Caprioli14a}. The corresponding energy spectra that characterize this turbulent state are likewise shown in Fig. \ref{fig:FraqCS}(Right).

We may therefore conclude that many nonlinear plasma configurations long suggested as sites of particle energization—such as large-scale current sheet or turbulent shocks—can, within only a few Alfvén times, transform into a similar multiscale environment filled with numerous coherent structures that exhibit characteristic power-law statistics.

\vspace{0.7cm}
\begin{center}
\fcolorbox{red}{red!10}{\parbox{8cm}{\color{blue}
From the analysis presented in this section, several main conclusions can be drawn:
\begin{itemize}

\item A defining property of strongly turbulent MHD flows is the spontaneous formation of coherent structures with characteristic power-law statistics.
\item In regimes where the fluid pressure greatly exceeds the magnetic pressure, these structures largely resemble those found in conventional hydrodynamic turbulence.
\item By contrast, when magnetic pressure dominates, coherent structures acquire distinct properties, and the emergence of current filaments, together with their mutual interactions, becomes a principal process shaping the evolution of the cascade.
\end{itemize}}}
\end{center}

\section{Selected Laboratory, Space, and Astrophysical Environments}

In laboratory, heliospheric, magnetospheric, and astrophysical settings, turbulence can be driven by a wide variety of physical processes---including pressure gradients, convection, shear, shocks, magnetic buoyancy, rotation, and gravity---yet it often evolves toward a qualitatively similar state: a multiscale, intermittent regime populated by coherent structures. The examples discussed below are therefore not intended to be exhaustive. Rather, they are selected to illustrate characteristic environments in which the progression from turbulent driving to structure formation is especially transparent.

\subsection{\bf Laboratory plasmas: edge turbulence in tokamaks}

\begin{figure}[!ht]
    \centering
\includegraphics[width=0.4\linewidth]{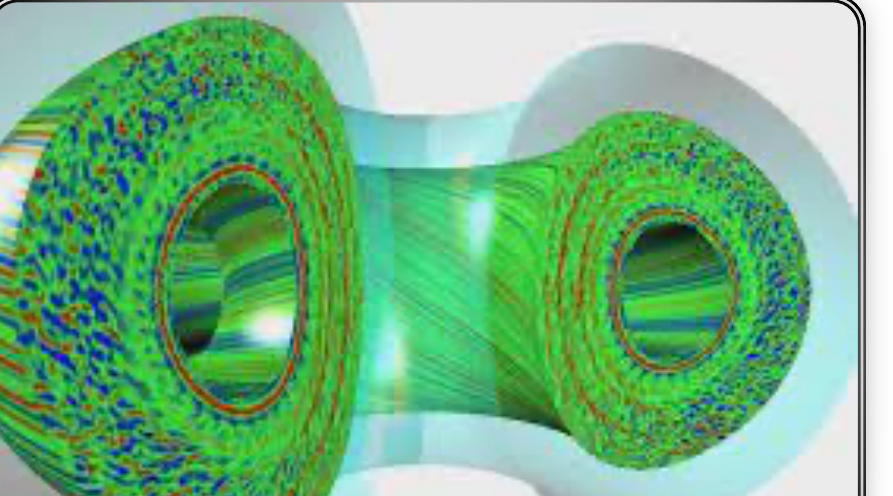}
    \caption{Visualization of microturbulence data generated by the Gyrokinetic Toroidal Code (GTC) \citep{PPPL2004GTCviz}.}
    \label{fig:toka}
\end{figure}

In tokamaks, edge turbulence arises naturally from the strong density, temperature, and pressure gradients that form near the separatrix and across the scrape-off layer. In this boundary region, the combined influence of unfavorable magnetic curvature, finite collisionality, and parallel losses along open field lines destabilizes the plasma through drift-wave, interchange, and ballooning-like modes. Nonlinear coupling among these instabilities, mediated by fluctuating $\mathbf{E}\times\mathbf{B}$ advection, generates broadband turbulence, intermittent cross-field transport, and filamentary coherent structures commonly known as blobs (Fig.~\ref{fig:toka}).

These blobs play a major role in regulating the particle and heat fluxes reaching the wall and, in turn, in determining both the width and the dynamical character of the edge transport layer \citep{Isliker2022}. Even in improved-confinement regimes, where shear in the radial electric field can partially reduce turbulent transport, the plasma edge remains highly structured and dynamically complex.

\subsection{MHD turbulence in the heliosphere}

\paragraph{\bf Solar convection zone.}
The solar convection zone is a strongly stratified and highly turbulent region in which convective motions over a broad range of scales continuously stretch, shear, fragment, and amplify magnetic fields.

\begin{figure}[!ht]
    \centering
    \includegraphics[width=0.4\linewidth]{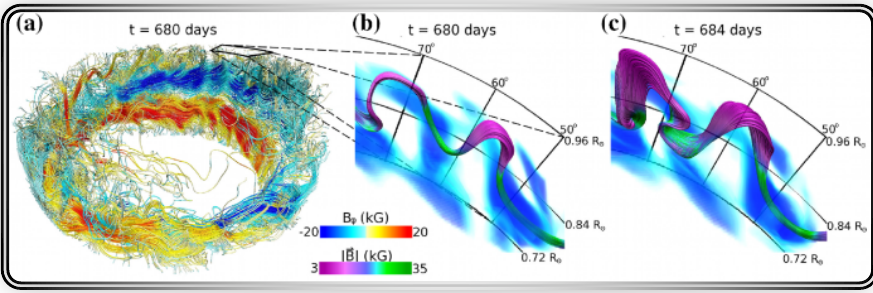}
    \caption{Within the convection zone, turbulent motions generate organized magnetic structures known as flux tubes, which rise toward the solar surface and eventually form active regions in the solar atmosphere \citep{Fan2009}.}
    \label{fig:convtur}
\end{figure}

Differential rotation, together with dynamo action, generates large-scale toroidal magnetic configurations, some of which become buoyantly unstable and reorganize into coherent magnetic flux tubes (Fig. \ref{fig:convtur}). The subsequent ascent of these tubes is shaped by turbulent buffeting, drag, rotation, and magnetic twist; taken together, these effects determine whether the tubes preserve their coherence and ultimately emerge at the solar surface as active regions. Turbulence in the convection zone thus plays a dual role: it generates magnetic intermittency while also allowing selected magnetic structures to survive as rising, organized flux systems.

\begin{figure}[!ht]
    \centering
\includegraphics[width=0.15\linewidth]{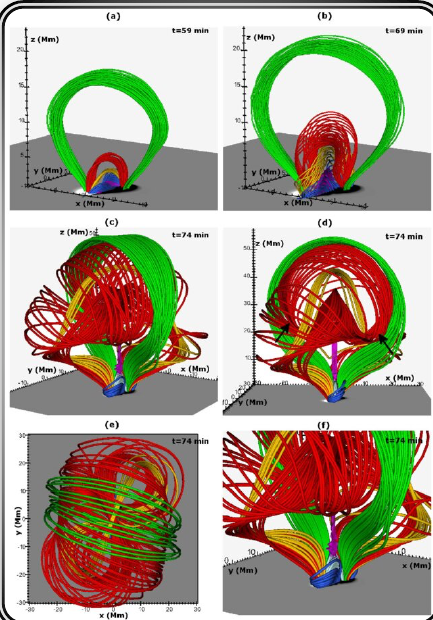}
\includegraphics[width=0.4\linewidth]{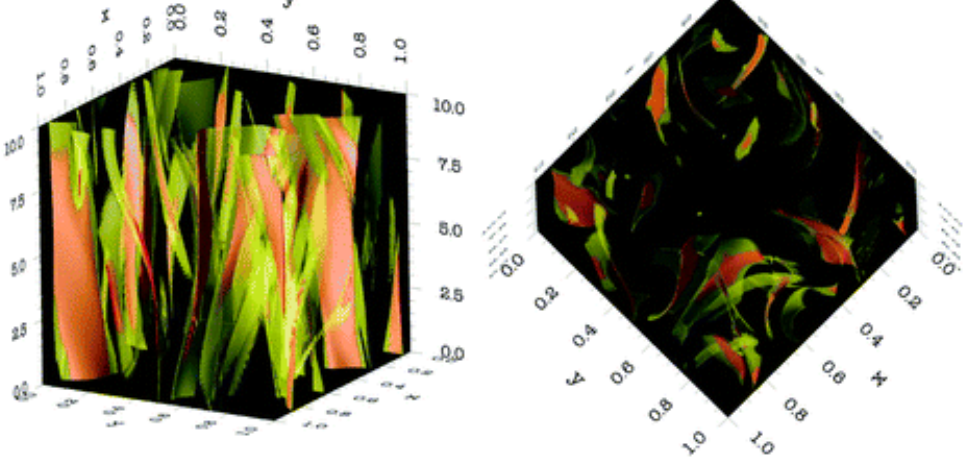}
    \caption{(Left) Persistent stressing of the footpoints of the emerging magnetic field generates coherent structures inside the active region, which subsequently power heating and flare-associated processes across a broad spectrum of spatial scales \citep{Archontis2010}. (Right) \citet{Rappazzo08} show that current sheets can form dynamically inside a coronal loop.}
    \label{fig:solaratm}
\end{figure}

\paragraph{\bf Solar atmosphere.}
The solar atmosphere, extending from the photosphere to the corona, behaves as a continuously driven magnetized system in which footpoint motions, emerging magnetic flux, propagating waves, and reconnection events inject fluctuations over a wide range of spatial and temporal scales (Fig. \ref{fig:solaratm}a).
As these perturbations propagate upward into layers where the density decreases sharply and the magnetic field becomes increasingly structured, partial wave reflection and nonlinear interactions promote Alfv\'enic turbulence, while field-line braiding and current-sheet formation localize dissipation in space and time (Fig. \ref{fig:solaratm}b).

\paragraph{\bf Solar wind.}
The solar wind provides the most accessible natural laboratory for the study of MHD turbulence, carrying from the Sun a broad spectrum of fluctuations that evolve as the flow expands through interplanetary space. On large scales, especially within high-speed streams, these fluctuations are often strongly Alfv\'enic.
\begin{figure}[!ht]
    \centering
\includegraphics[width=0.3\linewidth]{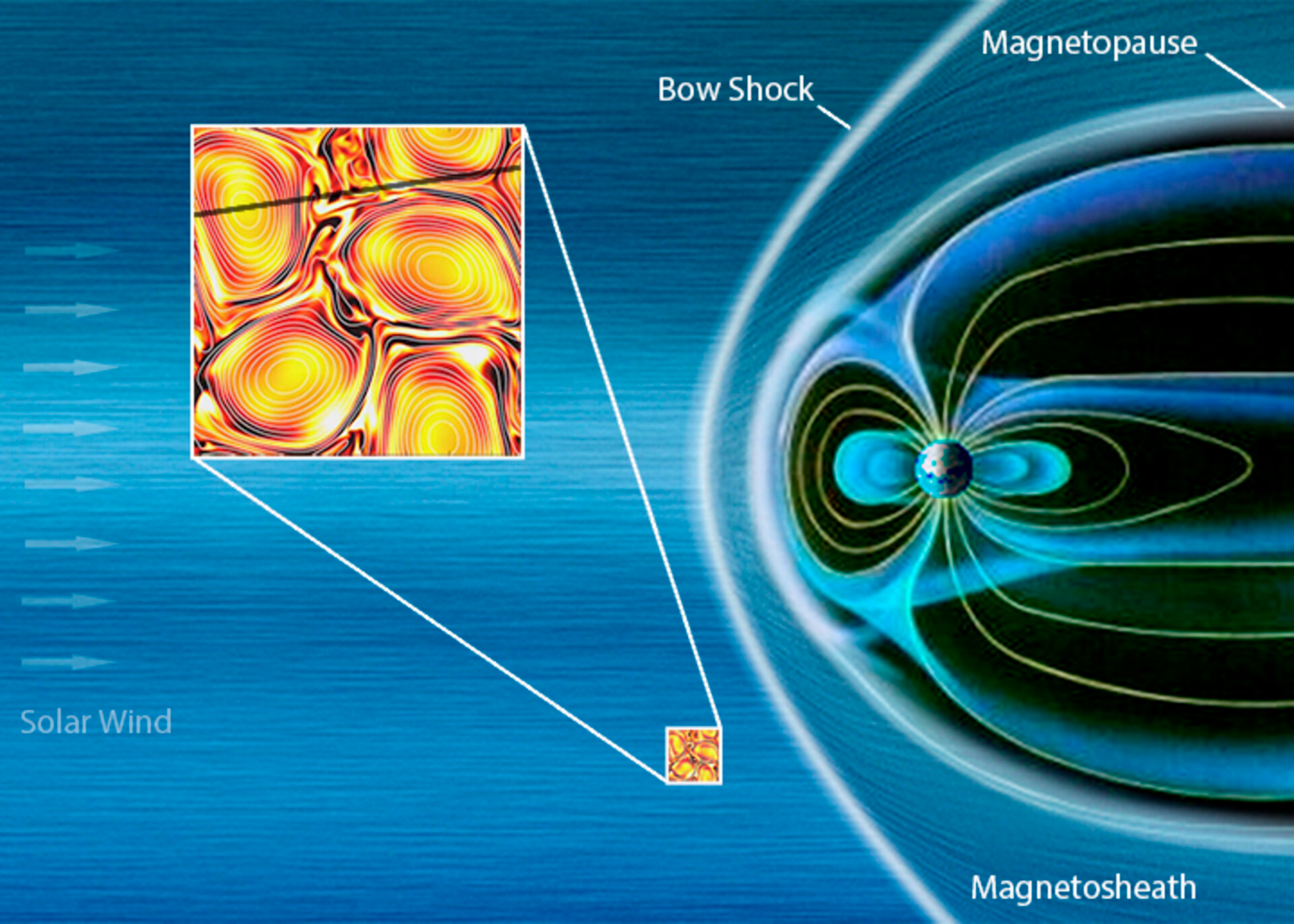}
\includegraphics[width=0.4\linewidth]{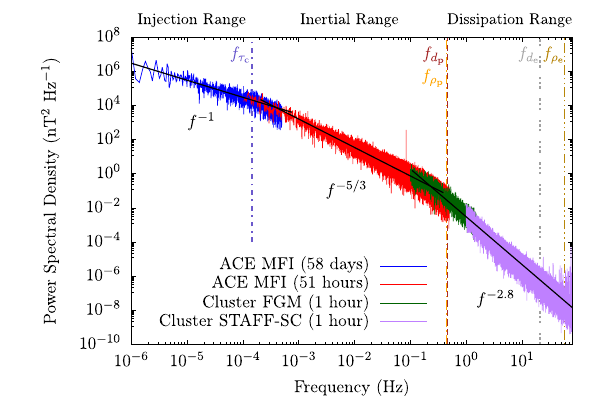}
    \caption{(Left) Turbulence in the solar wind. (Right) Power spectral density of magnetic field fluctuations in the solar wind. The black lines show power-law fits with slopes of $-1$, $-5/3$, and $-2.8$, corresponding to the injection, inertial, and dissipation ranges, respectively \citep{Verscharen19}.}
    \label{fig:solarwind}
\end{figure}

With increasing heliocentric distance, however, they develop stronger anisotropy and intermittency, accompanied by a cascade that transfers energy down to ion and electron kinetic scales (Fig. \ref{fig:solarwind}). Closer to the Sun, reflection-driven interactions between outward- and inward-propagating Alfv\'enic wave packets are thought to be particularly important, linking coronal dynamics to solar-wind heating and acceleration, as well as to the formation of coherent structures such as current sheets and flux ropes.

\begin{figure}[!ht]
    \centering
    \includegraphics[width=0.4\linewidth]{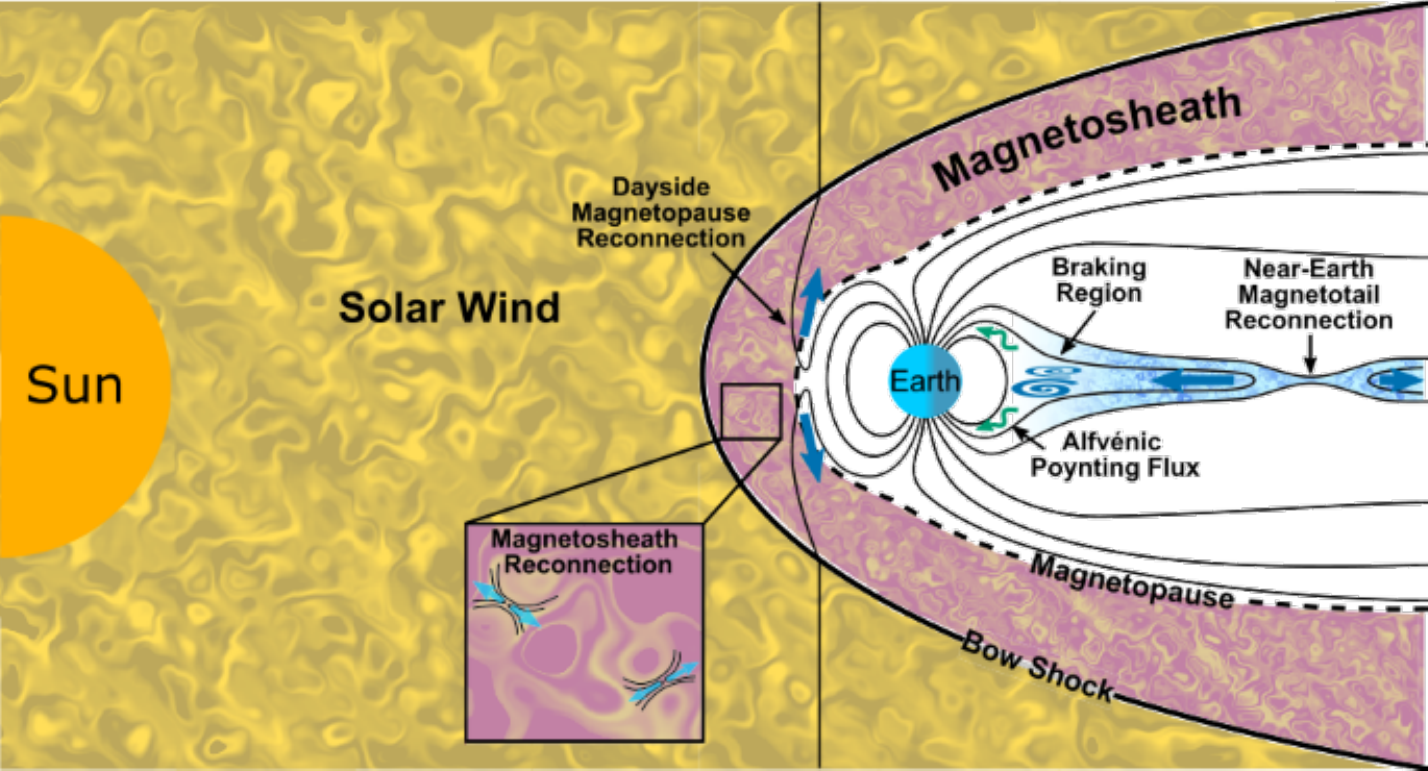}
    \caption{Diagram illustrating turbulent regions in near-Earth space together with the locations of large-scale reconnection events across the system. Turbulence can both interact with these global reconnection processes and generate additional, smaller-scale reconnection events within the turbulent regions \citep{Stawarz24}.}
    \label{fig:BowMag}
\end{figure}

\paragraph{\bf Bow shock and magnetosheath.}
At Earth's bow shock, the supersonic solar wind is abruptly slowed, compressed, and heated, creating a downstream magnetosheath filled with waves, discontinuities, and turbulent fluctuations. Rather than simply transmitting upstream turbulence unchanged, the shock transforms it through compression, unstable particle distributions, temperature anisotropies, and localized flow patterns, producing a highly intermittent plasma environment in which coherent structures and localized dissipation are widespread (Fig. \ref{fig:BowMag}). The magnetosheath therefore serves as an especially important transition region, linking large-scale shock-driven processes to reconnection, heating, and kinetic-scale dissipation.

\paragraph{\bf Earth's magnetotail.}
Earth's magnetotail is a driven and highly dynamic plasma environment in which energy supplied by the solar wind is first stored in elongated magnetic field lines and later released explosively through reconnection, bursty bulk flows, and current-sheet disruptions. These processes generate multiscale fluctuations in the magnetic field and plasma velocity within the plasma sheet, where vortical motions, filamentary current systems, and intermittent reconnection jets naturally produce a turbulent MHD state. The magnetotail is therefore not only a site of global magnetic reconfiguration, but also a region where turbulence, coherent structures, and reconnection are tightly interconnected.

\subsection{Selected astrophysical systems}

\begin{figure}[!ht]
    \centering
\includegraphics[width=0.25\linewidth]{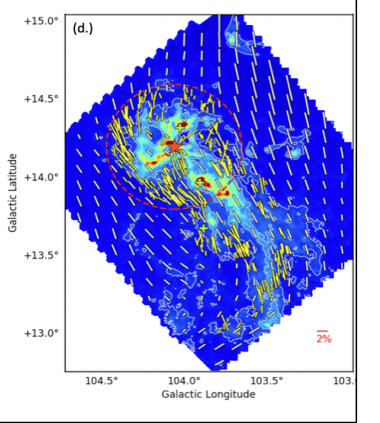}
    \caption{Magnetic-field structure in the bright-rimmed cloud complex L1172/1174. The background image traces the morphology of the cloud, while the short line segments indicate the projected magnetic-field orientation inferred from polarimetric measurements. The field exhibits a coherent, large-scale configuration, with localized bends and irregularities in denser regions, suggesting an interaction between the cloud structure, compressive processes, and magnetic-field topology during star formation \citep{Soam2024}.}
    \label{fig:starformation}
\end{figure}

\paragraph{\bf Star formation.}
Star formation takes place within cold, magnetized molecular clouds whose evolution is governed by highly compressible, often supersonic turbulence.

On large scales, this turbulence can oppose or reshape gravitational collapse, whereas on smaller scales it generates shocks, converging flows, and filamentary networks that lead to the formation of prestellar cores. Turbulence is therefore not a secondary complication superimposed on gravity, but a central process regulating fragmentation, star-formation efficiency, and the spatial clustering of young stellar populations (see Fig. \ref{fig:starformation})
.

\paragraph{\bf Accretion disks and the vicinity of black holes.}
Accretion disks around compact objects are intrinsically turbulent because microscopic viscosity alone is far too weak to account for the outward transport of angular momentum required for accretion.

\begin{figure}[!ht]
    \centering
\includegraphics[width=0.3\linewidth]{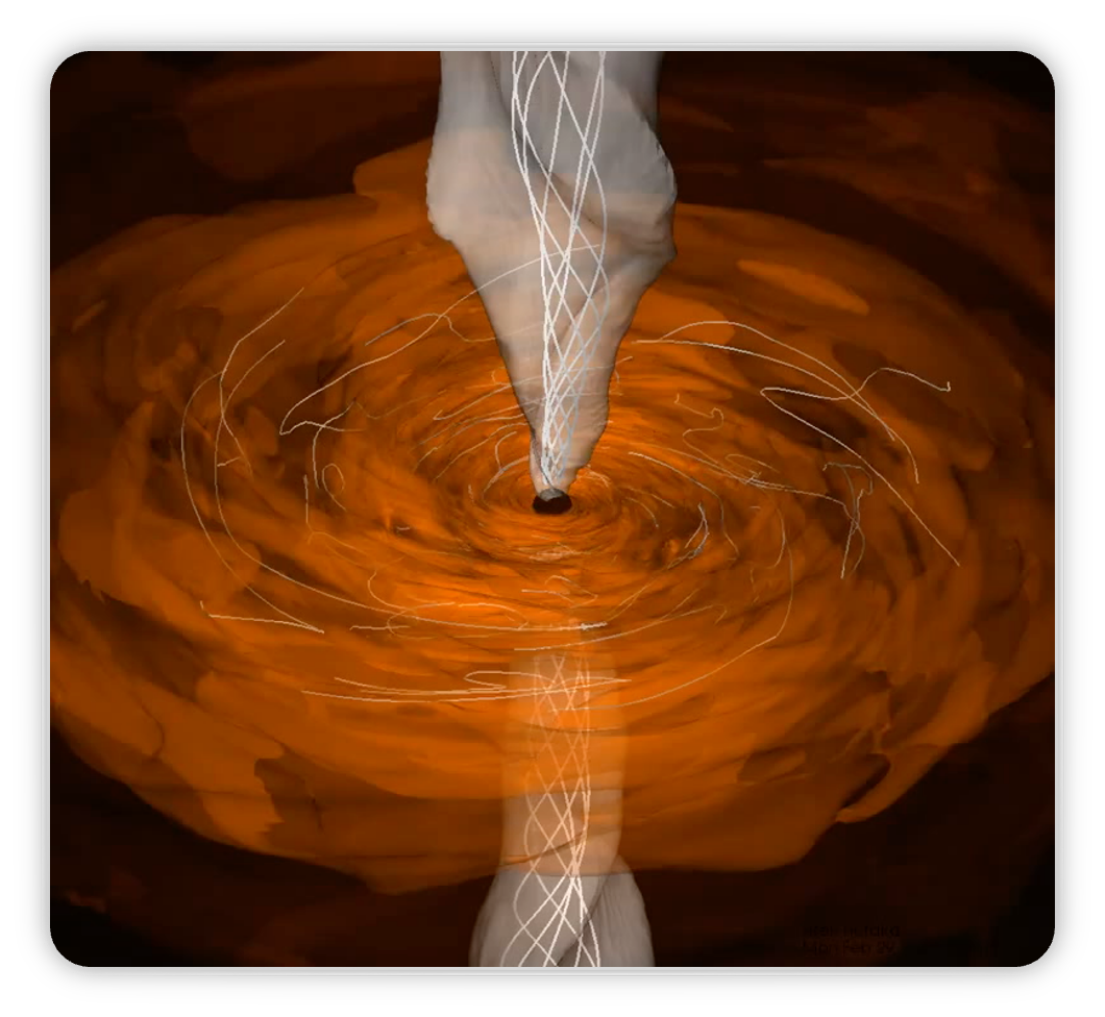}
    \caption{Schematic depiction of a magnetized accretion disk that feeds a compact central object and launches bipolar jets. The disk shows differential rotation, turbulent motion, and outward transport of angular momentum, while large-scale magnetic fields collimate and power outflows along the polar axis. Together, these interconnected multiscale plasma processes regulate energy dissipation and non-thermal particle acceleration in accreting systems.}
    \label{fig:jet}
\end{figure}

In sufficiently ionized disks, even relatively weak magnetic fields destabilize the shear flow through the magnetorotational instability, rapidly producing MHD turbulence that taps the free energy of orbital motion and converts it into stresses, dissipation, and enhanced accretion. In the vicinity of black holes, this turbulent plasma also interacts with large-scale magnetic fields, coupling the internal disk dynamics to the launching of relativistic jets and the production of high-energy emission. Turbulence in accretion flows is therefore more than a transport mechanism; it is a central process linking disk structure, variability, magnetic-energy release, and outflow formation.

\paragraph{\bf Astrophysical jets.}
Astrophysical jets, spanning systems from young stellar objects to active galactic nuclei and gamma-ray bursts, are highly collimated outflows whose evolution is strongly influenced by magnetized turbulence.
\begin{figure}[!ht]
    \centering
\includegraphics[width=0.3\linewidth]{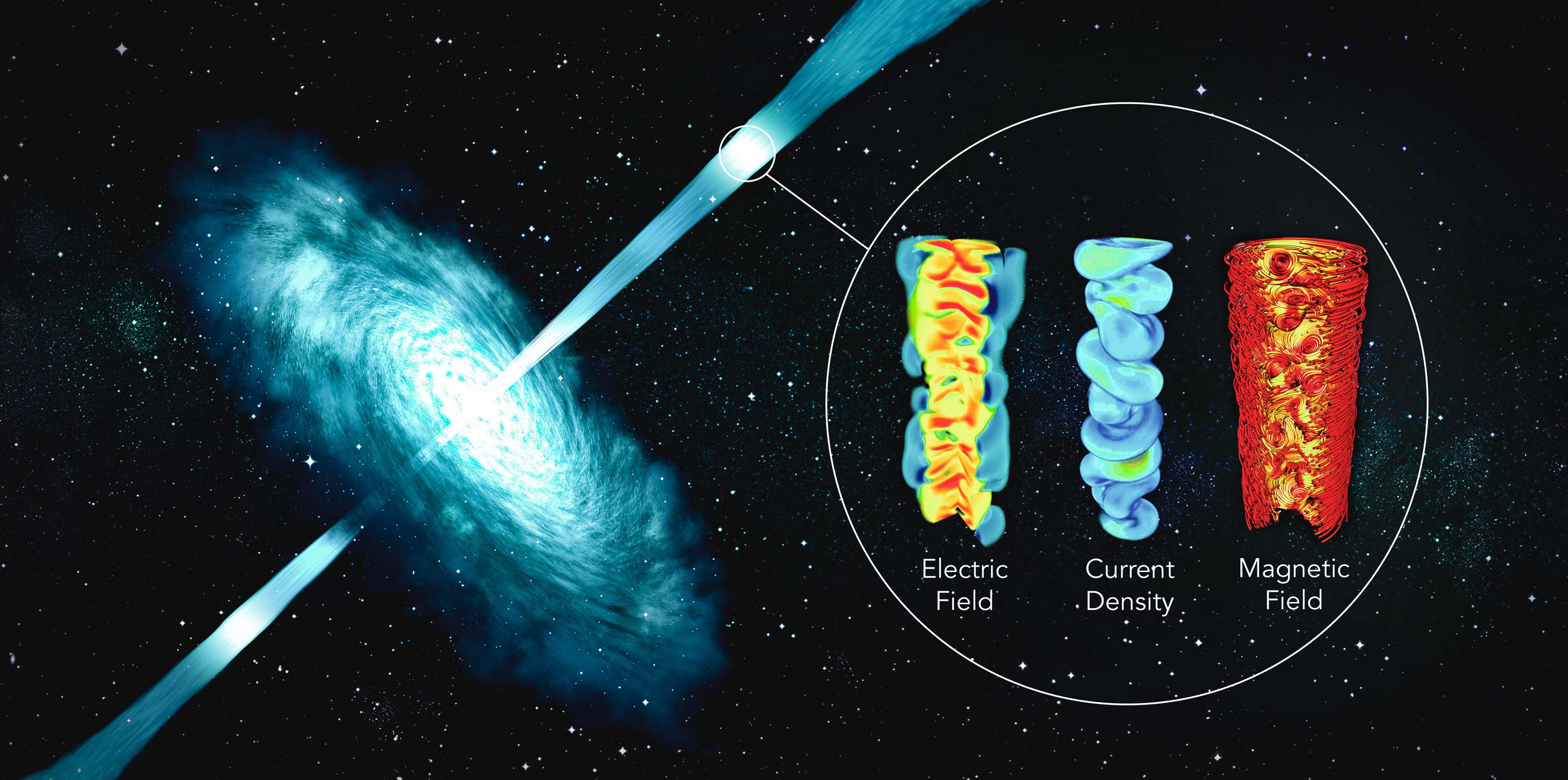}
    \caption{Artist's illustration of an astrophysical jet, with an inset displaying the jet's internal electromagnetic structure. The presence of irregular electric fields, localized current filaments, and a twisted (helical) magnetic field configuration indicates that relativistic jets provide favorable conditions for magnetic-energy dissipation, turbulent energy transport, and the acceleration of non-thermal particles \citep{alves18}.}
    \label{fig:jet2}
\end{figure}

After their launch, these jets propagate through stratified and inhomogeneous environments where shear, entrainment, current-driven instabilities, and internal shocks produce multiscale structure and turbulent mixing. This turbulence influences both jet stability and radiative output by enhancing magnetic dissipation, driving particle acceleration, and mediating the interaction with the surrounding medium. It is therefore essential not only for the persistence and collimation of jets, but also for their efficiency as sources of non-thermal radiation (see Fig. \ref{fig:jet2}).

\begin{figure}[!ht]
    \centering
\includegraphics[width=0.3\linewidth]{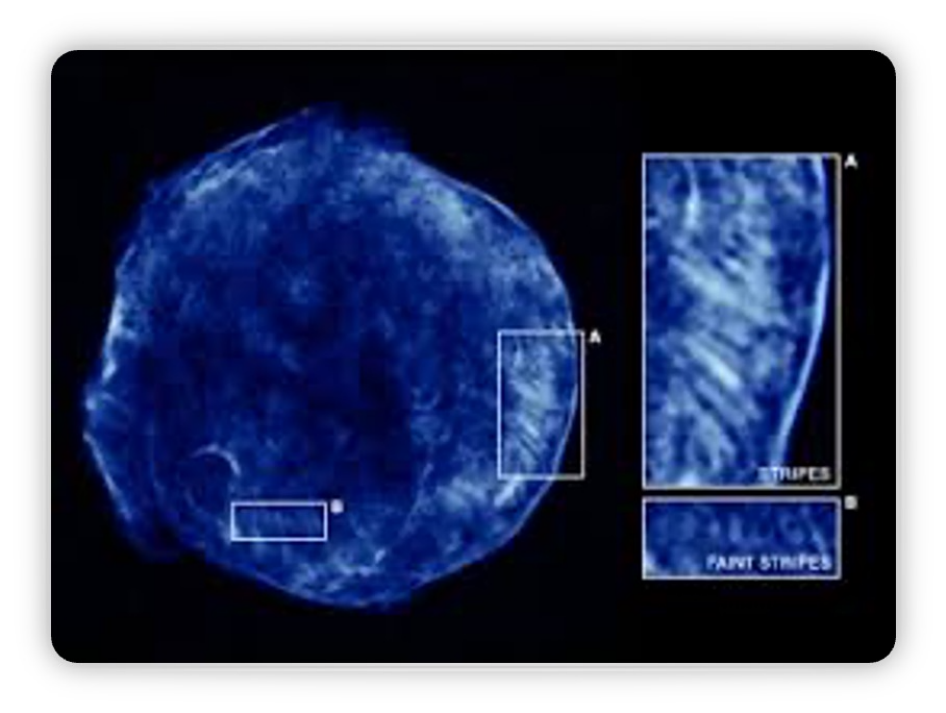}
\includegraphics[width=0.2\linewidth]{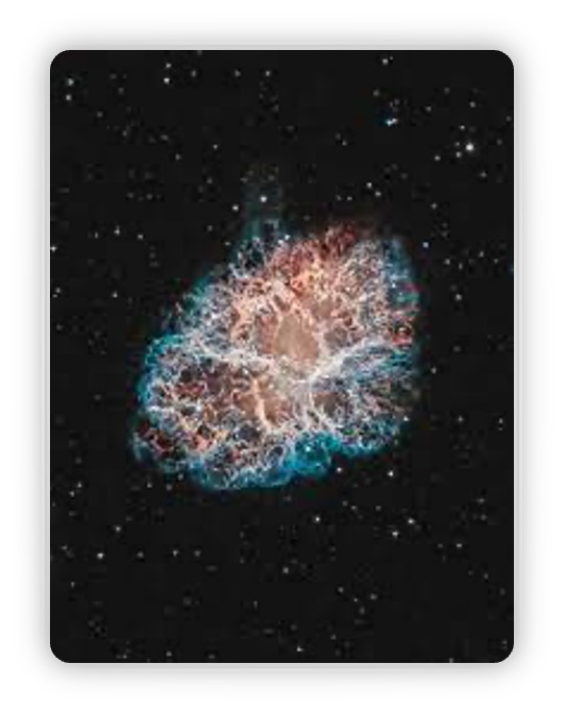}
    \caption{(Left) Young supernova remnant displaying narrow striped patterns and filamentary structures associated with turbulence driven by shocks and amplified magnetic fields. (Right) Supernova remnant showing an intricate network of luminous filaments formed by the nonlinear interaction between the outward-moving shock front and the surrounding plasma. These structures highlight the intermittent character of supernova remnants and their role as natural laboratories for the study of particle acceleration.}
    \label{fig:snr}
\end{figure}

\paragraph{\bf Supernova remnants.}
When a massive star ends its life in a supernova explosion, the blast wave drives an expanding remnant whose strong shocks sweep through the surrounding medium, creating a turbulent, magnetized shell (see Fig. \ref{fig:snr}a).

Turbulence develops both upstream and downstream of the shock, driven by shock corrugation, fluid instabilities, and the amplification of magnetic irregularities by cosmic rays. The result is a highly inhomogeneous and intermittent environment, widely regarded as one of the primary sites of Galactic cosmic-ray acceleration (see Fig. \ref{fig:snr}b). Supernova remnants therefore provide a particularly clear example of how shocks, turbulence, and magnetic-field amplification work together to convert explosion energy into heat, radiation, and non-thermal particles.

\begin{figure}[!ht]
    \centering
    \includegraphics[width=0.4\linewidth]{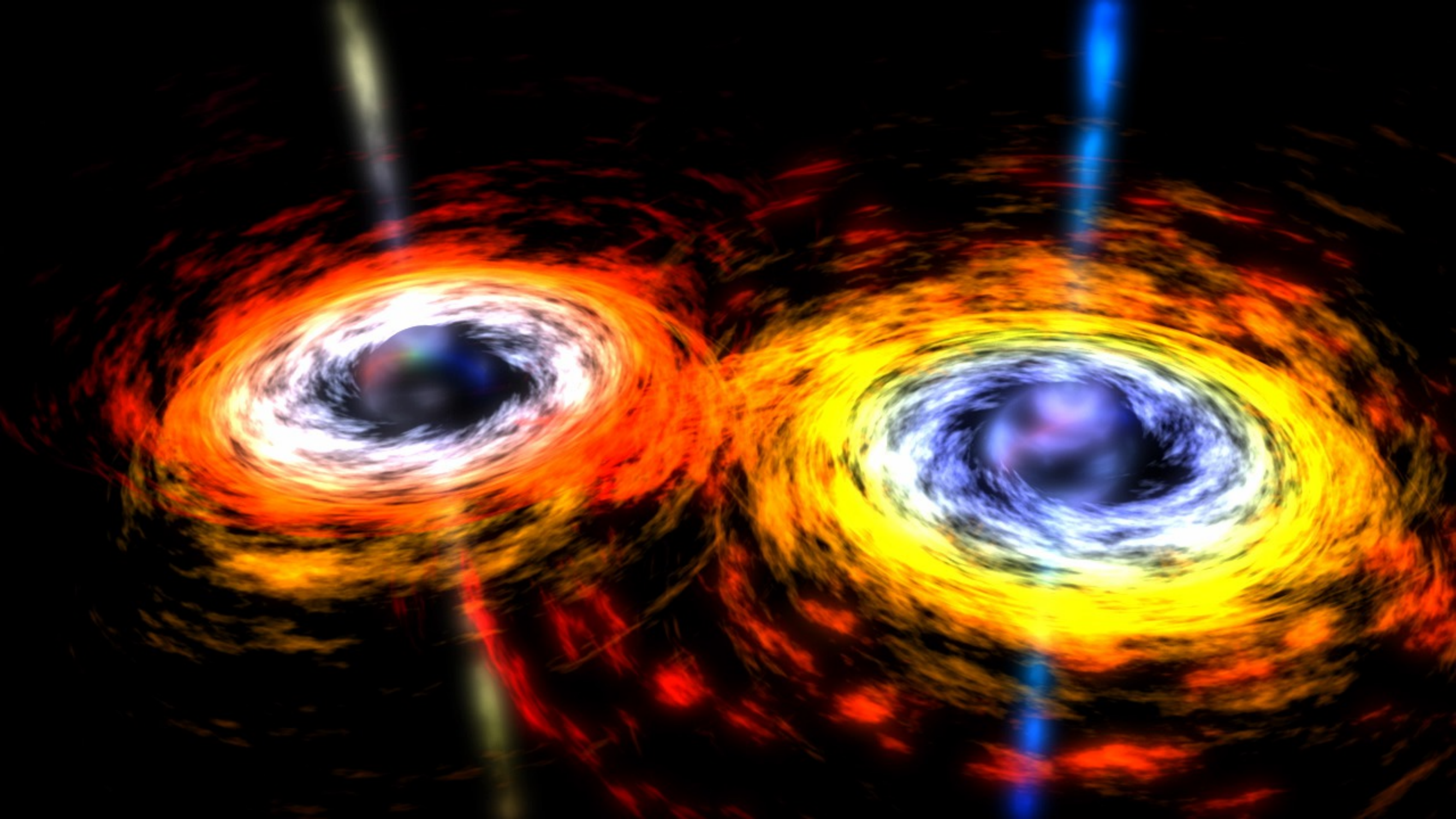}
    \caption{Artist's illustration of a pair of black holes in orbit, surrounded by bright accretion flows. Gas orbiting each compact object forms hot, chaotic accretion disks, while part of the infalling plasma is ejected as narrow relativistic jets aligned with the spin axes. These systems provide an extreme setting for angular-momentum transport, magnetic-energy dissipation, and the acceleration of high-energy particles.}
    \label{fig:blackholemerg}
\end{figure}

\paragraph{\bf Black-hole mergers and multimessenger environments.}
Mergers of compact objects have opened the era of multimessenger astronomy by connecting gravitational-wave detections with the electromagnetic response of the surrounding plasma and magnetic fields. In gas-rich environments, turbulence in circumbinary material, remnant accretion disks, or magnetized plasma following the merger can regulate angular-momentum transport, magnetic-field amplification, energy dissipation, and the possible formation of transient jets. Although these environments remain less well characterized than the cases discussed above, they may be especially important because the presence and nature of turbulence could determine whether a given merger is observed solely through gravitational waves or also through an electromagnetic counterpart (see Fig. \ref{fig:blackholemerg}).

\vspace{0.3cm}
\begin{center}
\fcolorbox{red}{red!10}{\parbox{0.8\linewidth}{\color{blue}
Although the systems discussed in this section are driven by very different physical mechanisms, they display a common phenomenology: turbulence generates multiscale coherent structures that regulate transport, dissipation, and, in many cases, particle energization. The recurrence of this behavior in laboratory, heliospheric, magnetospheric, and astrophysical plasmas supports the view that strong turbulence is ubiquitous in the universe and should be regarded as a dynamical state that naturally gives rise to structure.
}}
\end{center}
\section{Turbulence-Induced Coherent Structures as Particle Accelerators}

In strongly turbulent media, charged particles interact continuously with multiscale electromagnetic fluctuations as well as with coherent structures generated by the turbulent cascade, such as current sheets, vortices, shocks, contracting magnetic islands, and localized reconnection regions \citep{Fermi1949, Fermi1954, Isliker17a, Lemoine2021}. These interactions collectively produce a combination of stochastic energization, systematic acceleration, particle trapping and escape, and complex transport in both configuration and energy space \citep{Teaca2014, Isliker17a, Lemoine2021, Pezzi22}.

\subsection{Cosmic rays and non-thermal particles in the heliosphere}

\begin{figure}[!ht]
    \centering
\includegraphics[width=0.3\linewidth]{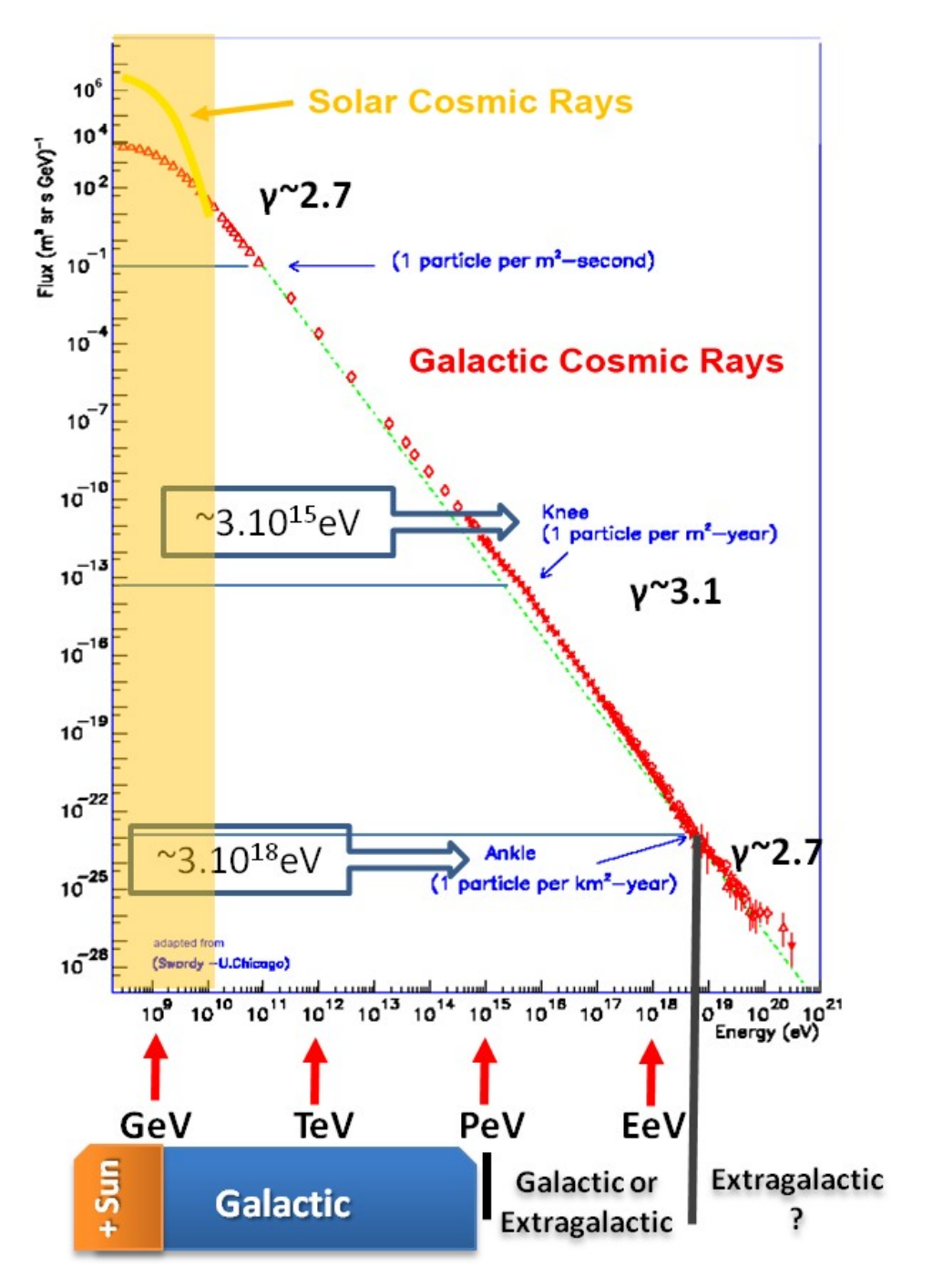}
    \caption{Measured all-particle cosmic-ray energy spectrum, highlighting the solar contribution at low energies and the principal spectral features of the Galactic and extragalactic components, in particular the knee at about $3\times10^{15}\,\mathrm{eV}$ and the ankle near $3\times10^{18}\,\mathrm{eV}$ \citep{PDG2024CosmicRays}.}
    \label{fig:CRspec}
\end{figure}

Cosmic rays constitute the non-thermal component of astrophysical plasmas and provide one of the clearest indications that efficient particle acceleration operates far from thermal equilibrium. Their nearly ubiquitous power-law energy spectrum points to robust acceleration processes active across a wide variety of cosmic environments, while the almost isotropic flux measured at Earth reflects the dominant role of magnetic scattering during propagation through interstellar and heliospheric magnetic fields. At the same time, small but measurable deviations from isotropy persist, especially at the highest energies, indicating that acceleration and transport remain closely coupled (see Fig. \ref{fig:CRspec}).

Supernova-remnant shocks are still widely regarded as the primary mechanism responsible for producing Galactic cosmic rays up to at least the knee \citep{Hillas2005,Schure2012}. Yet cosmic rays are far more than passive tracers of explosive activity. Through pressure feedback, ionization, plasma heating, and the conditions under which they are produced, they actively modify the dynamics of their surroundings. They therefore represent a major channel through which turbulent and explosive plasma flows transfer and redistribute energy over a broad range of scales.

\begin{figure}[!ht]
    \centering
\includegraphics[width=0.3\linewidth]{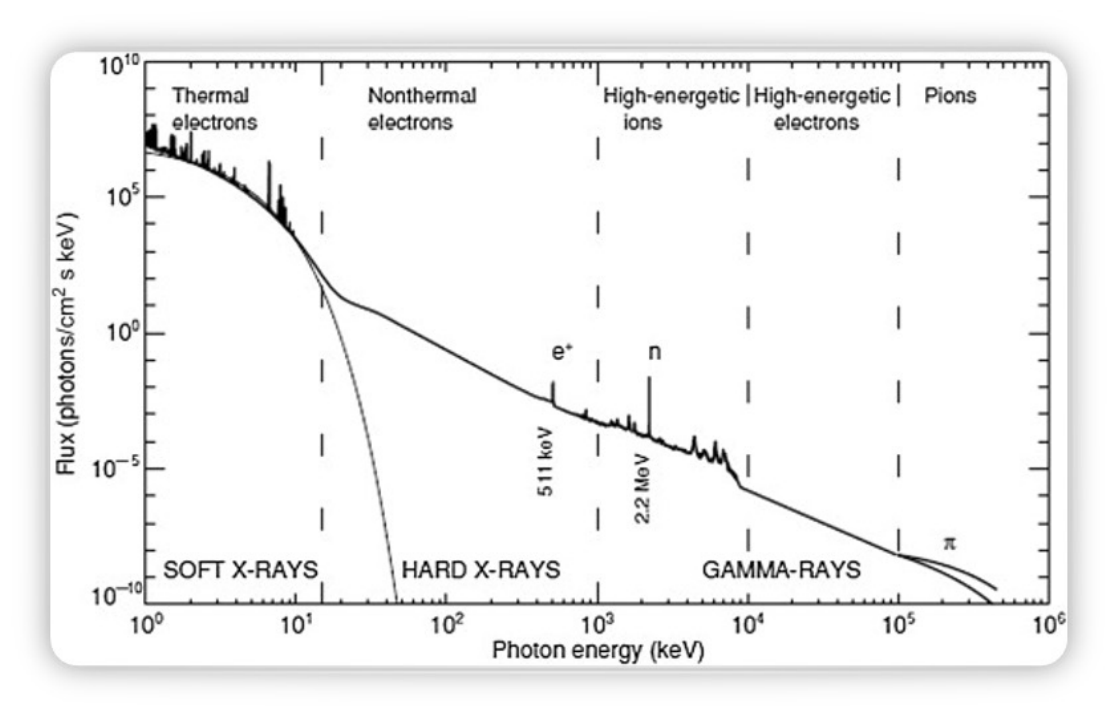}
\includegraphics[width=0.3\linewidth]{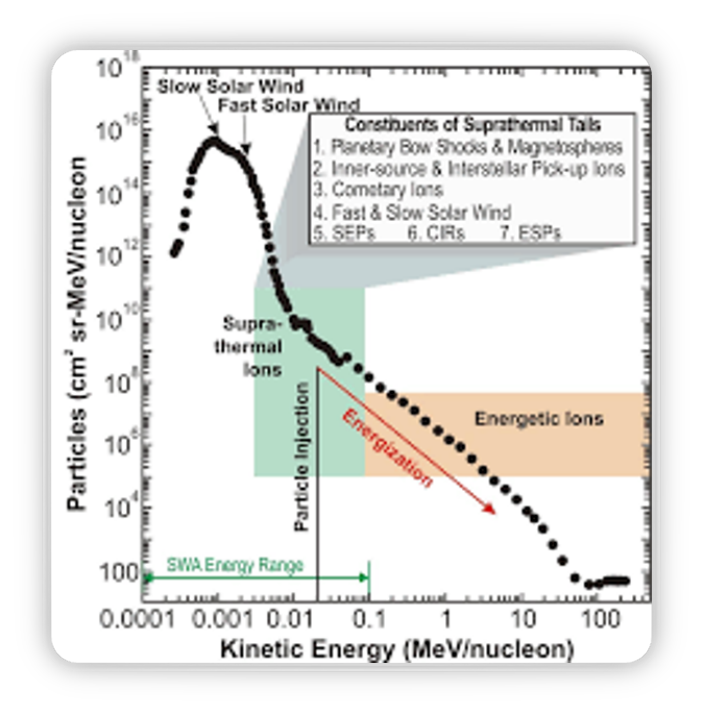}
    \caption{(Left) Broadband photon spectrum of a major solar flare, showing the transition from thermal soft X-ray emission to non-thermal hard X-rays and gamma rays produced by accelerated electrons and ions, as well as by nuclear interactions and pion decay \citep{Lin03}. (Right) Energy spectrum of heliospheric ions extending from the solar-wind peak through the suprathermal tail into the energetic-particle regime. The shaded transition region marks the injection window, where seed particles are supplied for further acceleration by shocks and turbulent structures in the heliosphere \citep{Mewaldt2001}.}
    \label{fig:flareswaccel}
\end{figure}

The heliosphere provides a particularly instructive example of this broader picture. In the low corona, impulsive magnetic-energy release, magnetic reconnection, and turbulent dissipation accelerate both electrons and ions, while CME-driven shocks produce gradual solar energetic-particle events (see Fig. \ref{fig:flareswaccel}a). As the solar wind expands outward, additional energization occurs through wave--particle interactions, turbulent scattering, magnetic islands, and reconnection within heliospheric current sheets and stream-interaction regions. Taken together, these processes sustain suprathermal particle populations and supply seed particles for subsequent shock acceleration \citep{Reames1999,Vlahos2019,Phan2022}. Particle energization at the Sun and throughout the solar wind should therefore be viewed as a tightly coupled multiscale process linking coronal dynamics, turbulence, reconnection, and shocks (see Fig. \ref{fig:flareswaccel}b).

The broader conclusion is that particle heating and non-thermal acceleration in the heliosphere are not distinct phenomena, but different manifestations of the same turbulent hierarchy that also characterizes many of the energetic systems discussed earlier.

\subsection{From Fermi acceleration to coherent-structure acceleration}

\begin{figure}[!ht]
    \centering
    \includegraphics[width=0.4\linewidth]{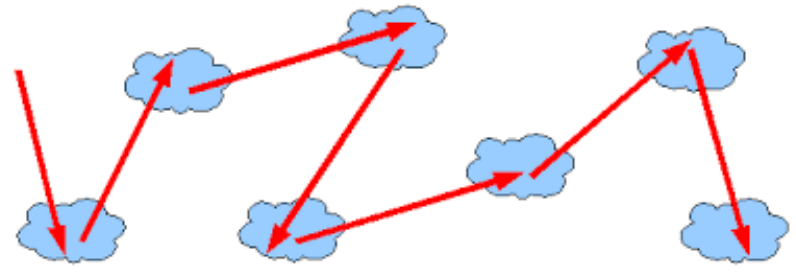}
\includegraphics[width=0.2\linewidth]{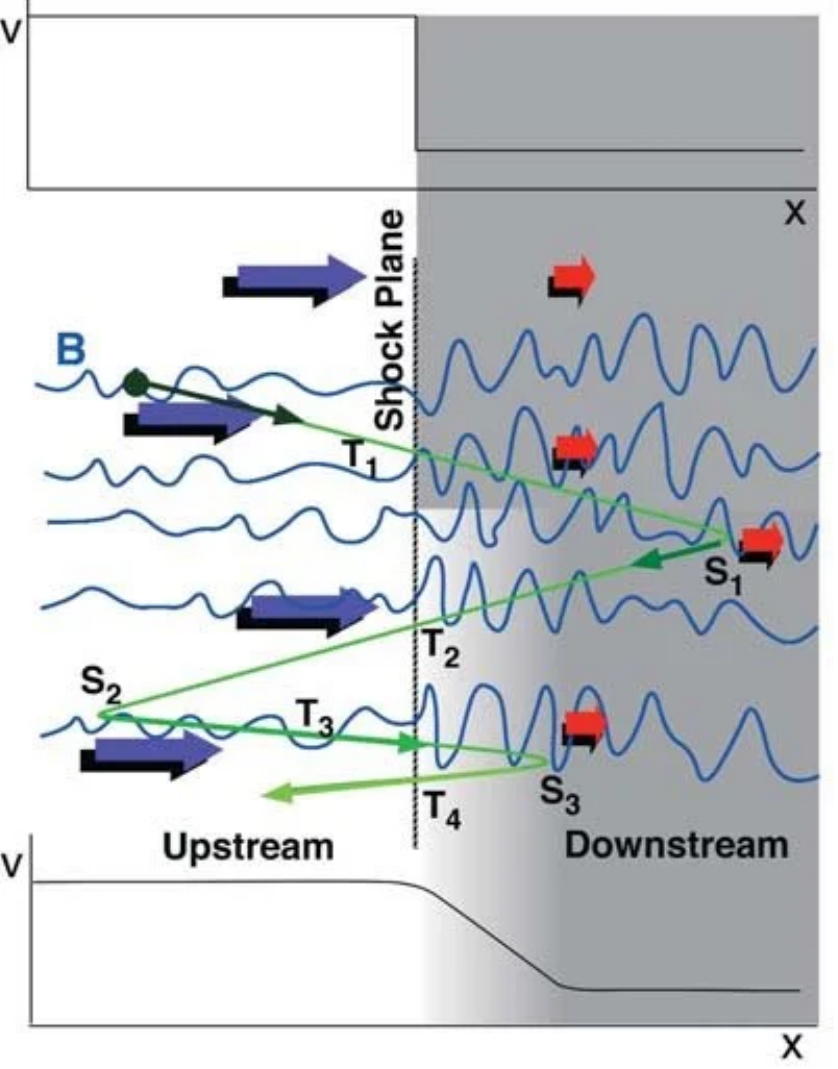}
    \caption{(Left) {\bf Stochastic} (second-order) Fermi acceleration results from random interactions with moving magnetic irregularities or scattering centers. (Right) First-order Fermi acceleration occurs at a shock front, where repeated crossings produce a {\bf systematic} increase in particle energy.}
    \label{fig:SecFirstFermi}
\end{figure}

The modern discussion of particle acceleration begins with Fermi's seminal proposal that repeated interactions between fast particles and moving magnetic irregularities can produce net energization \citep{Fermi1949}. In its original form, this stochastic process is second-order in the ratio $V/c$, where $V$ is the speed of the scattering centers, and is therefore relatively inefficient (see Fig. \ref{fig:SecFirstFermi}Left). Fermi soon recognized that converging flows, and especially shock fronts, provide a much more efficient first-order mechanism \citep{Fermi54} (see Fig. \ref{fig:SecFirstFermi}Right). This insight led to the theory of diffusive shock acceleration, which remains one of the cornerstones of high-energy astrophysics \citep{Longair11}.

\begin{figure}[!ht]
    \centering
\includegraphics[width=0.6\linewidth]{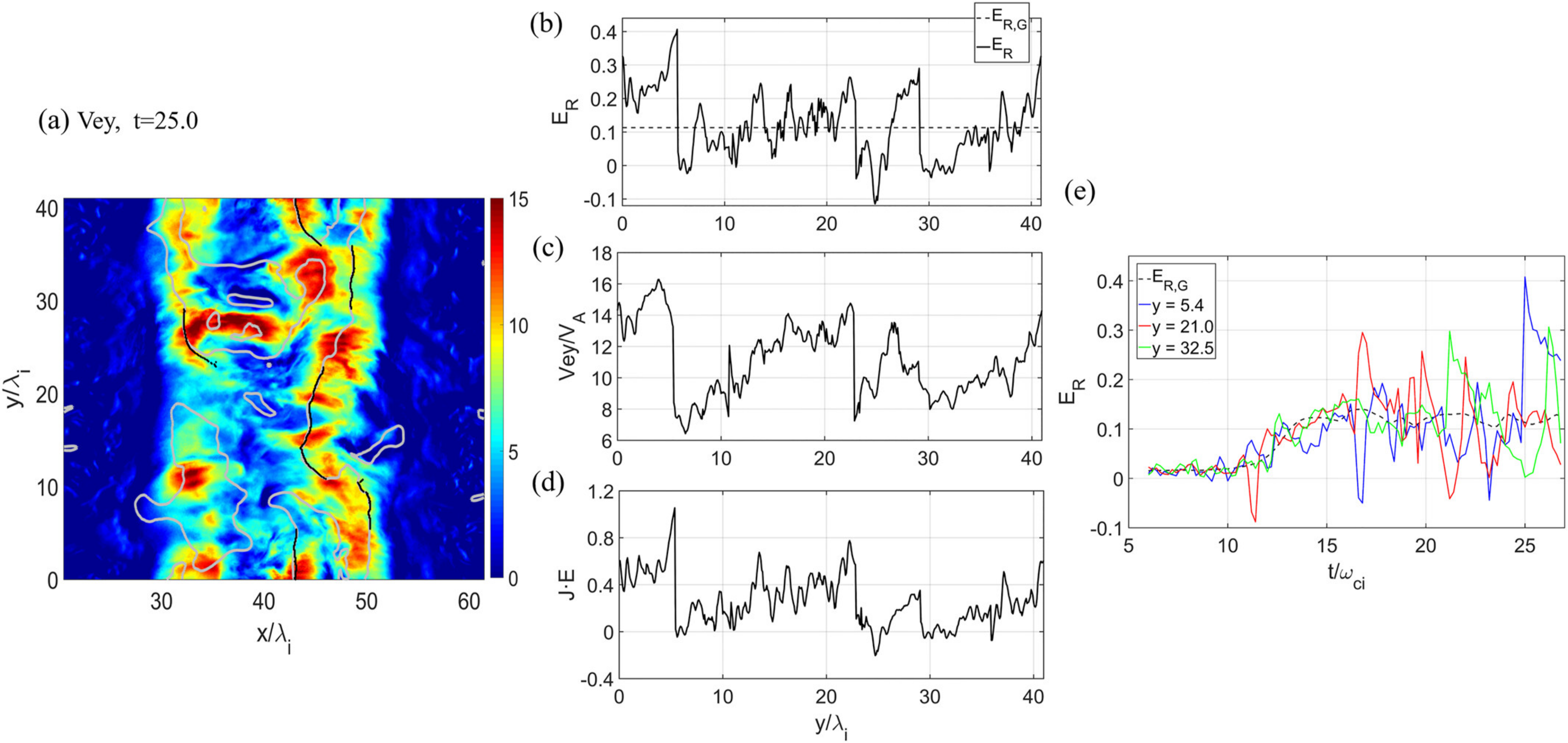}
    \caption{(a) Evolving turbulent current sheet. (b)--(d) Spatial profiles of (b) $E_R$, (c) $V_{ey}$, and (d) $\vec{J} \cdot \vec{E}$ along the most active x-line, with the dashed curve indicating the global reconnection rate. (e) Temporal evolution of $E_R$ \citep{Liu24}. Particle motion in the evolving turbulent electric fields leads to heating and acceleration within the current sheet.}
    \label{fig:CSacc}
\end{figure}

Several years after Fermi first introduced these ideas, reconnecting current sheets were identified as another fundamental and persistently active acceleration site, especially in heliospheric and magnetospheric environments, where they were associated with solar flares and Earth's magnetotail \citep{deJager1986,LitvinenkoSomov1993,LitvinenkoCraig1996,Litvinenko97} (see Fig. \ref{fig:CSacc}). Stochastic Fermi acceleration, shock acceleration, and acceleration in reconnecting current sheets have therefore come to be regarded as the standard mechanisms responsible for producing non-thermal particles in space and astrophysical plasmas.

The perspective adopted here does not replace these mechanisms. Rather, it embeds them within a broader turbulent framework. In strongly turbulent plasmas, different coherent structures coexist and interact with particles along their trajectories. Since Ohm's law is
\begin{equation}
\mathbf{E}=-\frac{\mathbf{u}\times\mathbf{B}}{c}+\eta\mathbf{J},
\end{equation}
the fluctuating velocity field naturally produces stochastic energization through the $-\mathbf{u}\times \mathbf{B}/c$ term, while intense localized current structures generate coherent electric-field components through $\eta \mathbf{J}$ that are responsible for first-order acceleration (see \citet{Isliker25, Lemoine25} for details). Strong turbulence therefore couples heating and acceleration in a persistent way, naturally producing particle distributions with a heated core and a suprathermal or non-thermal tail (Kappa distribution).

\subsection{MHD turbulence and test particles}

A widely used approach evolves the turbulent plasma through the MHD equations and then follows charged particles as test particles in the resulting electromagnetic fields \citep{Onofri06,Turkmani05,Isliker19, Teaca2014,Lemoine2021}. In this framework, particle feedback on the fields is neglected, an assumption that remains valid as long as the energy density of the accelerated particles is small compared with the magnetic and bulk-flow energy densities.

A particle of charge $q$ and mass $m$ obeys the relativistic Lorentz force,
\begin{equation}
\frac{d\mathbf{x}}{dt}=\mathbf{v},
\qquad
\frac{d\mathbf{p}}{dt}=q\left(\mathbf{E}+\frac{\mathbf{v}\times\mathbf{B}}{c}\right),
\qquad
\mathbf{p}=\gamma m\mathbf{v},
\end{equation}
with energy evolution
\begin{equation}
\frac{d}{dt}(\gamma mc^2)=q\,\mathbf{v}\cdot\mathbf{E}.
\label{eq:energygain}
\end{equation}
Equation (\ref{eq:energygain}) makes the central point explicit: a particle's energy can change only through the electric field. The magnetic field redirects particle motion, controls trapping and confinement, and organizes transport, but it does not itself do work.

In ideal MHD, the electric field in the fluid frame is perpendicular to the magnetic field. Particles can nevertheless be energized through drifts in strongly nonuniform fields, as well as through compressive and time-dependent processes and the contraction of flux ropes, where strong inductive electric fields arise \citep{Lemoine23}. Test-particle simulations are therefore especially useful because they retain the multiscale structure and intermittency of turbulent MHD fields while allowing direct examination of particle orbits, trapping, escape, and energy change.

\begin{figure}[!ht]
    \centering
    \includegraphics[width=0.4\linewidth]{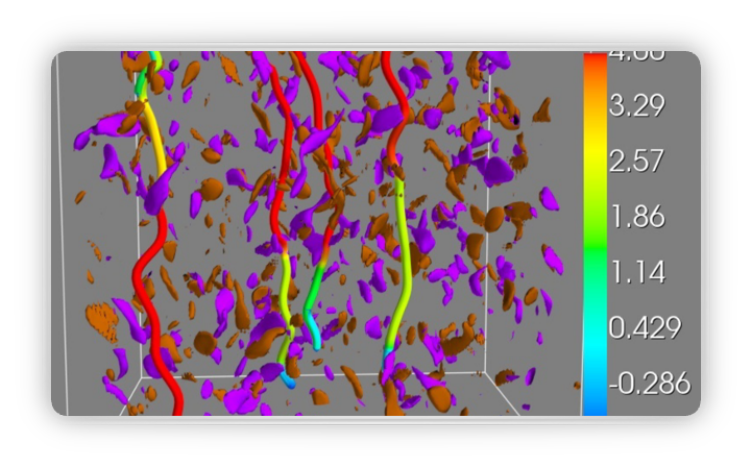}
\includegraphics[width=0.4\linewidth]{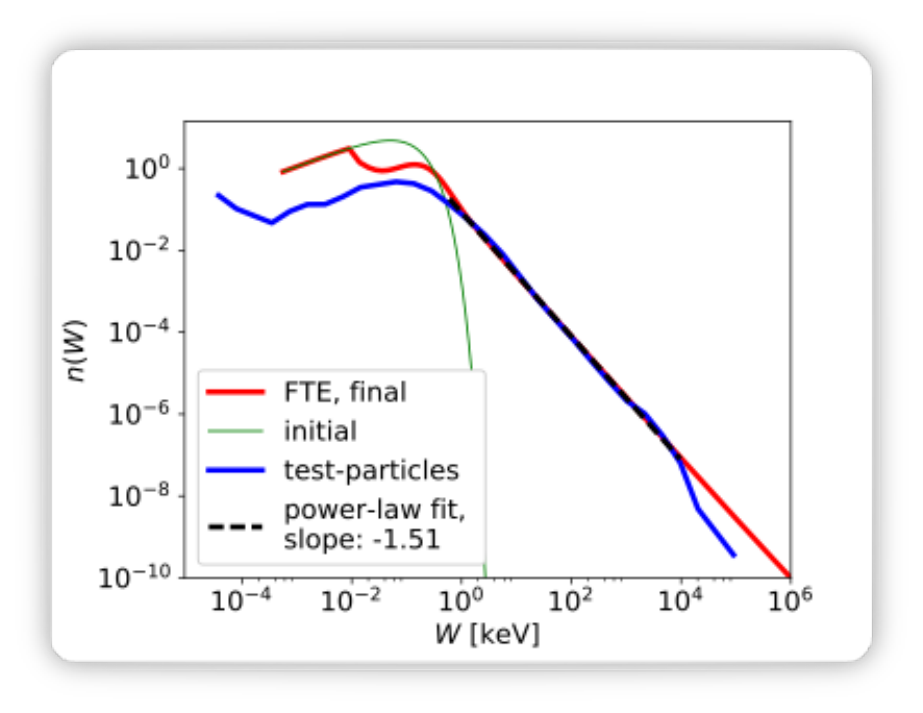}
    \caption{Representative results from test-particle calculations in strongly turbulent electromagnetic fields, illustrating particle transport and energization in an environment dominated by spontaneously emerging coherent structures.}
    \label{fig:MHDtest}
\end{figure}

\citet{Isliker17a} analyzed particle transport in strongly developed turbulence, where the random spatial distribution of spontaneously emerging coherent structures dominates the energy gain (Fig.~\ref{fig:MHDtest}).

At the ensemble level, the particle distribution function is often described by a Fokker--Planck equation,
\begin{align}
\frac{\partial P(W,t)}{\partial t}
=
-\frac{\partial [F(W)P(W,t)]}{\partial W}
&+ \frac{\partial^2 [D(W)P(W,t)]}{\partial W^2}\nonumber\\
&-
\frac{P(W,t)}{t_{\rm esc}(W)},
\label{FokkerPVelLoss}
\end{align}
where $F(W)$ and $D(W)$ describe systematic and stochastic energy changes, respectively, and $t_{\rm esc}(W)$ denotes the escape time from a bounded system \citep{Schhlickeiser02Book}. This framework is appropriate when particles undergo many small, weakly correlated interactions. By contrast, in strongly intermittent turbulence, broad distributions of both waiting times and energy increments may emerge---especially when coherent structures control the dynamics---so that conventional diffusive transport may need to be generalized to fractional or event-driven formalisms \citep{Isliker25, Effenberger25}.

The strength of the test-particle method lies in its intermediate character. It preserves the multiscale structure of turbulent fields and shows that random scattering and systematic energy gain can operate simultaneously. Its limitation is equally clear: once energetic particles begin to affect the global dynamics appreciably, or when kinetic-scale physics becomes essential, a fully kinetic description is required.

\subsection{Particle-in-cell simulations}

Particle-in-cell (PIC) simulations provide the fully kinetic framework required in regimes where microscopic plasma processes cannot be neglected \citep{Dawson1983,Pritchett2003,BirdsallLangdon1991,HockneyEastwood1988}. In the PIC method, ions and electrons are represented by computational macro-particles whose trajectories are advanced self-consistently in electromagnetic fields defined on a discrete grid, while those fields are updated through Maxwell's equations using the charge and current carried by the particles.

The main advantage of PIC lies in its ability to resolve kinetic phenomena absent from MHD and test-particle approaches: charge separation, plasma oscillations, kinetic instabilities, electron and ion diffusion regions, non-Maxwellian velocity distributions, and the self-consistent feedback of particles on the electromagnetic fields. This is crucial for studies of particle acceleration because reconnection layers, shock injection, and the most intense dissipation regions are often governed by kinetic-scale physics.

A major practical limitation is the separation of scales. To resolve kinetic physics, a PIC simulation must capture characteristic plasma scales such as the Debye length, the plasma frequency, and the skin depth \citep{Sturrock94}. This requirement is computationally demanding for astrophysical systems whose macroscopic dimensions exceed kinetic scales by many orders of magnitude. PIC, therefore, cannot yet simulate an entire turbulent corona, jet, or magnetotail with a realistic scale hierarchy. It should instead be viewed as a local microscope that reveals the elementary kinetic processes operating inside the dissipative structures produced by larger-scale turbulence.

\begin{figure}[!ht]
    \centering
\includegraphics[width=0.4\linewidth]{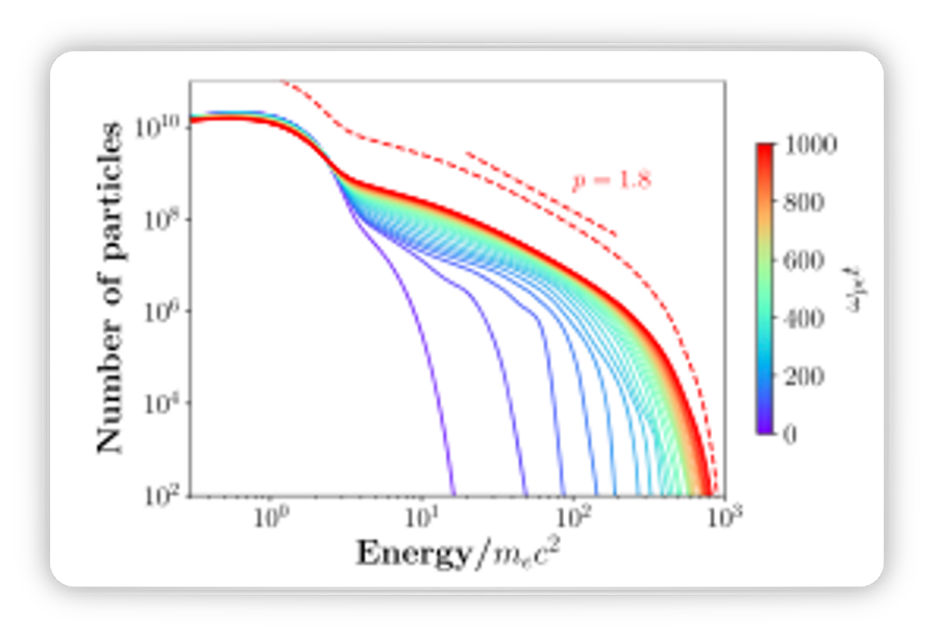}
    \caption{Particle energy spectra at different times during a representative PIC simulation. At late times, the spectrum develops a clear non-thermal tail with a spectral index of $p = 1.8$ \cite{Guo21}.}
    \label{fig:PICspec}
\end{figure}

\citet{Comisso19} explored strong turbulence and particle acceleration using PIC simulations, and \citet{Guo21} studied the temporal evolution of a reconnecting current sheet (see Fig. \ref{fig:PICspec}).

MHD and test-particle approaches indicate where energization regions are expected to appear and characterize how particles move through a turbulent multiscale environment. PIC simulations, in turn, reveal the dynamics within those regions once fluid descriptions become invalid. A complete theory of turbulent particle acceleration must incorporate both perspectives.

\subsection{A Network of Scatterers}

Coherent structures determine the specific sites where intermittent, event-driven particle energization takes place. This viewpoint suggests that energization is more naturally described as a sequence of discrete events than as a smooth, continuous diffusion process in a uniform background. Particles spend most of their time in a relatively quiescent medium and gain the bulk of their energy during brief encounters with current sheets, shocks, flux ropes, or similar structures. When such structures are arranged in a fractal or multifractal manner, the resulting distributions of encounter distances, waiting times, and energy increments can naturally produce anomalous transport.
\begin{figure}[!ht]
    \centering
    \includegraphics[width=0.3\linewidth]{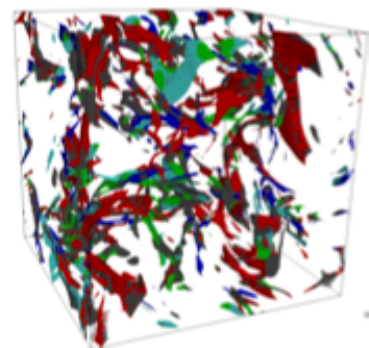}
\includegraphics[width=0.4\linewidth]{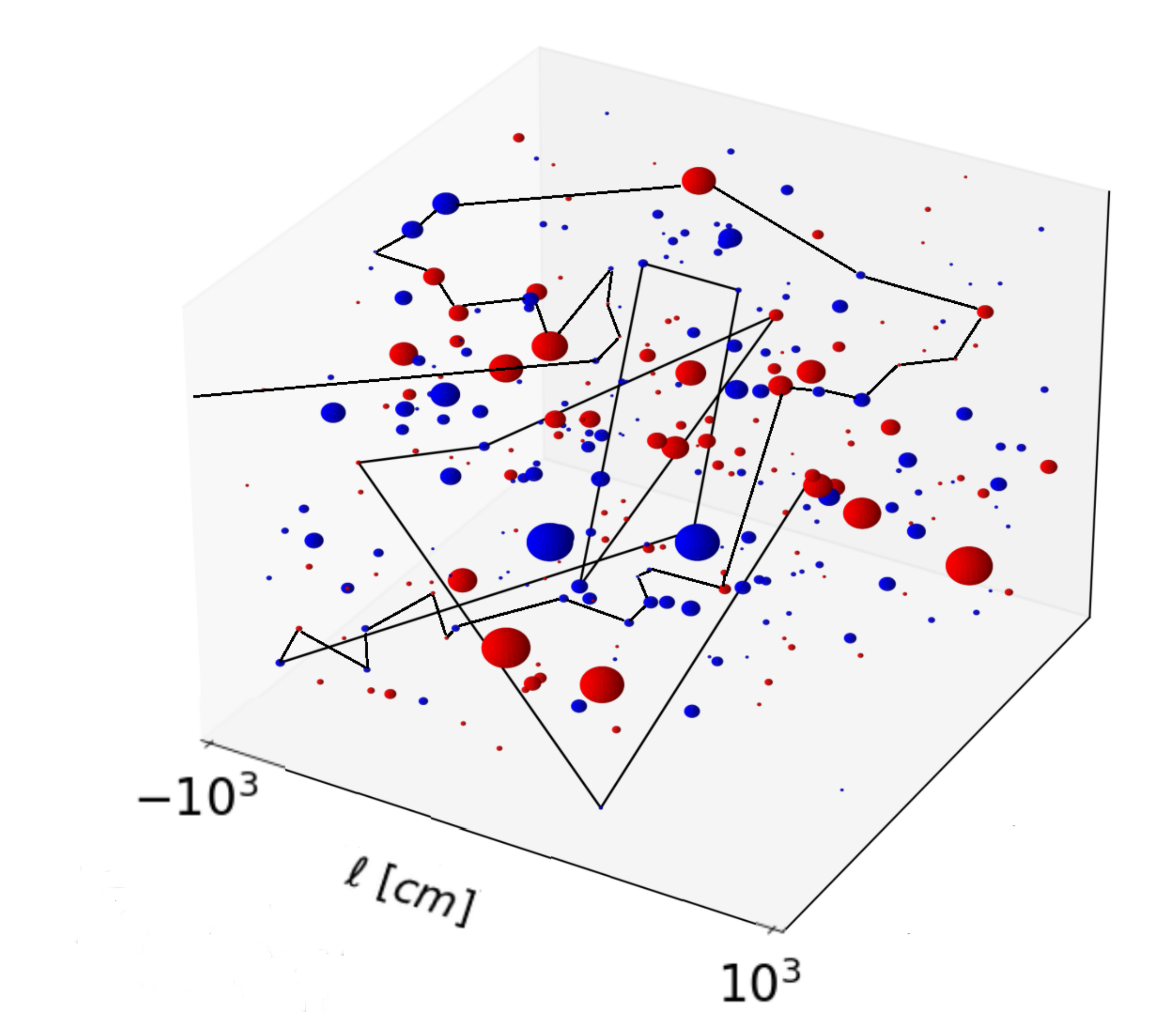}
    \caption{(Left) A range of coherent structures within a three-dimensional turbulent plasma \citep{Richard22}. (Right) Schematic depiction of coherent current structures in a magnified portion of the turbulent region. As particles move through this volume, they interact with these structures in both stochastic and systematic ways \citep{Sioulas22c}.}
    \label{fig:PINNS1}
\end{figure}

\begin{figure}[!ht]
    \centering
\includegraphics[width=0.25\linewidth]{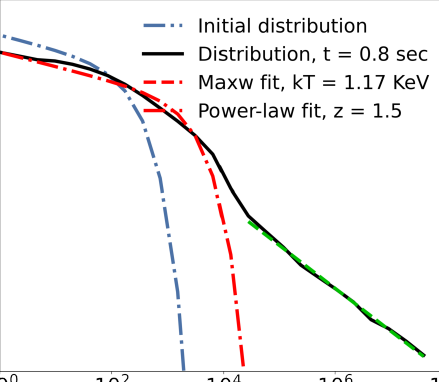}
    \caption{Steady-state kinetic-energy distribution showing the coexistence of a heated Maxwellian-like core and a power-law tail \citep{Sioulas22c}.}
    \label{fig:Sioulasf1}
\end{figure}

\citet{Sioulas22c} investigated this scenario in a highly turbulent, unbounded medium by representing coherent structures as a network of scatterers, each endowed with its own stochastic or systematic energization properties. In this framework, the key elements are the statistical distributions of free-path lengths and waiting times, along with the associated energy increments acquired at random. When waiting times are negligible and spatial displacements exhibit fractal characteristics, the particle distribution evolves into a heated core accompanied by a power-law tail (see Fig. \ref{fig:Sioulasf1}).

This is precisely the parameter regime in which reduced statistical descriptions and physics-informed surrogate models may become particularly useful. The aim of future Physics-Informed Neural Network (PINN) strategies is not to replace first-principles plasma physics, but to be guided by MHD and test-particle results while tracking the effective statistical behavior of energization events, waiting times, acceleration timescales, and escape times. In this way, one may eventually establish the missing feedback connection between large-scale turbulence and kinetic particle transport that is absent from the standard test-particle approximation.

\subsection{Toward reduced models and physics-informed surrogates}

One promising direction is the development of physics-informed surrogate models. In this context, the goal of a PINN-based or related framework is not to reproduce the full hierarchy of plasma dynamics through a purely data-driven fit. Rather, it is to infer reduced transport closures from simulation data while remaining consistent with physical constraints. The most appropriate targets are therefore not the full particle distribution function itself, but effective quantities such as systematic and stochastic energization coefficients, escape times, waiting-time distributions, or event-conditioned kernels (neural operators) that characterize particle interactions with coherent structures.

The motivation for such reduced models follows directly from the structure-centered picture advanced in this review. In a strongly turbulent plasma, particles do not gain energy uniformly throughout the volume. Instead, they spend much of their time in a comparatively quiet background and receive most of their energization during short encounters with current sheets, shocks, flux ropes, or other coherent structures. A useful coarse-grained description should therefore focus on the statistics of discrete energization events rather than assume spatially smooth diffusion from the outset.

A practical strategy would thus be to combine MHD simulations and test-particle calculations within an event-resolved framework. The MHD simulation provides the evolving electromagnetic fields together with a catalog of coherent structures. Test-particle trajectories then supply information on where and when particles encounter these structures, how long they remain trapped, and how much energy they gain or lose during each interaction. From these data, one may infer the distributions of energy increments, waiting times, and free-path lengths, as well as the distributions of acceleration and escape times in a finite turbulent domain.

Within this setting, physics-informed surrogates may serve as reduced-closure models. Their role would be to learn, from simulation data and explicit transport constraints, which effective description is appropriate under given conditions: a standard diffusive closure, a fractional transport model, or a more general event-based statistical description. This should still be viewed as a developing strategy rather than an established predictive framework. Its usefulness will depend critically on the quality of the structure diagnostics, the definition of the events, and the physical constraints built into the surrogate.

The value of such an approach, if successful, would be modest but important. It could provide a principled way to connect large-scale turbulence, coherent-structure statistics, and reduced kinetic transport without requiring a fully kinetic simulation of the entire astrophysical system. In that sense, physics-informed surrogates may eventually provide part of the missing bridge between multiscale MHD turbulence and non-thermal particle evolution.

\section{Summary}

The central claim of this article is clear. Strong MHD turbulence should be viewed as a structure-forming dynamical state. Once this perspective is adopted, plasma heating, the emergence of suprathermal tails, and non-thermal particle acceleration can be interpreted within a single physical framework. From this standpoint, MHD turbulence is not simply compatible with cosmic particle acceleration; it is one of the most natural environments in which it arises.

\begin{center}
\fcolorbox{red}{red!10}{\parbox{9cm}{\color{blue}
A particularly important implication of the coherent-structure-based picture advanced here is that it may unify two of nature's most striking scale-free laws: the Kolmogorov spectrum of turbulence and the power-law energy distributions of high-energy particles and cosmic rays.}}
\end{center}



\end{document}